\documentclass{aa}  

\usepackage{graphicx}
\usepackage{txfonts}
\usepackage{subcaption}         
\usepackage{lscape}             
\usepackage{placeins}           
                                
\usepackage[dvipsnames]{xcolor}
\usepackage{adjustbox}
\usepackage{hyperref}
\usepackage{natbib}
\usepackage{makecell}
\usepackage{siunitx}
\usepackage{multirow}
\usepackage{pdflscape}
\usepackage{pgffor}
\usepackage{multicol}
\usepackage{rotating, threeparttable}
\usepackage{array}
\usepackage{float}
\usepackage[version=4]{mhchem}
\usepackage[normalem]{ulem}
\usepackage[morefloats=500]{morefloats}
\usepackage{xstring}
\newcommand{\plotfig}[4][0.85\linewidth]{%
  \StrSubstitute{#2_#3}{_}{-}[\safeLabel]%
  \begin{figure}[H]
    \centering
    \includegraphics[width=#1]{Figures/Line_profile_fits/#2_#3_line_spectra_fits.pdf}
    \caption{#4}
    \label{fig:\safeLabel}
  \end{figure}
}

\definecolor{Blue}{rgb}{0,0,1}
\hypersetup{colorlinks, citecolor=Blue, linkcolor=Blue, urlcolor=Blue}

\DeclareUnicodeCharacter{2212}{-}

\defcitealias{Unnikrishnan_et_al_2024}{Paper I}
\defcitealias{Unnikrishnan_et_al_2025}{Paper II}

\begin{document}

   \title{Charting circumstellar chemistry of carbon-rich \\asymptotic giant branch stars}

   \subtitle{III. SiO and SiS abundances}


   \author{R. Unnikrishnan
          \inst{1}
          \and
          E. De Beck\inst{1}
          \and
          L.-\AA. Nyman\inst{1,2,3}
          \and
          H. Olofsson\inst{1}
          \and
          W. H. T. Vlemmings\inst{1}
          \and
          M. Maercker\inst{1}
          \and
          M. Van de Sande\inst{4}
          \and
          T. J. Millar\inst{5}
          \and
          T. Danilovich\inst{6,7}
          \and
          M. Andriantsaralaza\inst{1}
          \and
          S. B. Charnley\inst{8}
          \and
          M. G. Rawlings\inst{9}
          }

   \institute{Department of Space, Earth and Environment, Chalmers University of Technology, {SE-412 96 Gothenburg}, Sweden\\\email{ramlal.unnikrishnan@chalmers.se}
      \and
      Joint ALMA Observatory (JAO), Alonso de Córdova 3107, Vitacura 763-0355, Casilla 19001, Santiago, Chile
      \and
      European Southern Observatory (ESO), Alonso de Córdova 3107, Vitacura 763-0355, Santiago, Chile
      \and
      Leiden Observatory, Leiden University, PO Box 9513, NL-2300 RA Leiden, The Netherlands
      \and
      Astrophysics Research Centre, School of Mathematics and Physics, Queen’s University Belfast, University Road, Belfast, BT7 1NN, UK
      \and
      School of Physics \& Astronomy, Monash University, Wellington Road, Clayton 3800, Victoria, Australia
      \and
      Institute of Astronomy, KU Leuven, Celestijnenlaan 200D, 3001 Leuven, Belgium
      \and
      NASA Goddard Space Flight Center, 8800 Greenbelt Road, Greenbelt, MD 20771, USA
      \and
      Gemini Observatory / NSF NOIRLab, 670 N. A’ohoku Place, Hilo, Hawai’i, 96720, USA
      }

   \date{Received 16 December 2025 / Accepted 6 April 2026}

  \abstract
   {The circumstellar envelopes of AGB stars are sites of rich molecular chemistry. The present understanding of C-rich AGB chemistry largely relies on observations of the archetypal carbon star IRC$+$10$\,$216. Current molecular abundance estimates for carbon stars are based either on single-dish spectra sampling a range of excitation conditions, or on interferometric mapping of a few lines.}
   {We aim to estimate the circumstellar abundances of SiO, SiS, and their {most abundant} isotopologues {($^{29}$SiO, $^{30}$SiO, $^{29}$SiS, $^{30}$SiS, and Si$^{34}$S)} for a sample of five carbon stars. This study compares molecular abundances across the sources, tests chemical modelling predictions, and examines whether IRC$+$10$\,$216 is representative of the broader carbon star population.} 
   {We derived molecular abundances using detailed 1D non-local thermodynamic equilibrium (non-LTE) radiative transfer (RT) modelling, constrained by both morphological and excitation information obtained from spatially resolved ALMA maps and single-dish observations. We further compared the derived abundances to chemical modelling results.}
   {We obtain good fits to the SiO and SiS line profiles, and derived well-constrained abundance profiles and reliable isotopic ratios for all sources except AFGL 3068. While the SiS peak abundances are very similar across the sample (2.0$\times$10$^{-6}-4.7\times$10$^{-6}$), we find that the SiO peak abundances of the rest of the stars are a factor of $\sim$5 larger than that of IRC~+10\,216. The $e$-folding radii ($R_\mathrm{e}$) are in the range 1.3$\times$10$^{16}-7.0\times$10$^{16}$ cm for SiO and 6.0$\times$10$^{15}-1.0\times$10$^{17}$ cm for SiS. The $R_\mathrm{e}$ increases with gas density for both SiO and SiS. Our RT models cannot simultaneously fit the low- and high-$J$ SiO lines of IRC+10216. Chemical models reproduce the derived SiO abundance profiles well, while over-predicting the SiS $R_\mathrm{e}$ values.}
   {Our models highlight the necessity of having spatially resolved observations across a broad range of excitation conditions to robustly constrain molecular abundance profiles, while also making evident the limitations inherent in 1D RT modelling using simplified (circum)stellar models. We find that the currently assumed SiS photodissociation rate in chemical models is underestimated.}


   \titlerunning{SiO and SiS abundances in carbon stars}
   \authorrunning{R. Unnikrishnan et al.}

    \maketitle
    \nolinenumbers

\section{Introduction}
The circumstellar envelopes (CSEs) of asymptotic giant branch (AGB) stars are rich in gas and dust, and provide ideal conditions for a wide range of astrochemical processes leading to the formation of numerous molecular species \citep[e.g.][]{Cernicharo_et_al_2000, Hofner_and_Olofsson_2018, Unnikrishnan_et_al_2024, Agundez_et_al_2020, Van_de_Sande_et_al_2019}. In carbon-rich (C-type) AGB CSEs, where the photospheric carbon-to-oxygen ratio (C/O) exceeds unity, complex molecules, including long carbon chains \citep[e.g.][]{Agundez_et_al_2017, Gong_et_al_2015, Woods_et_al_2003, Unnikrishnan_et_al_2024} and possibly also polycyclic aromatic hydrocarbons \citep[PAHs, e.g.][]{Cherchneff_2012, Tielens_2008, Zeichner_et_al_2023, Anand_et_al_2023} can form efficiently. {Molecules as large as C$_{60}$ and C$_{70}$ have been detected in planetary nebulae, which are the descendants of AGB CSEs \citep{Cami_et_al_2010}}. The extended outflows from these stars shape the chemical composition of the interstellar medium (ISM), enriching it with the raw material for future star and planet formation \citep[e.g.][]{Hofner_and_Olofsson_2018, Tielens_2005, Matsuura_et_al_2009, Kobayashi_et_al_2011}.

Several molecules commonly found in C-type AGB CSEs, such as CO, C\textsubscript{2}H\textsubscript{2}, CS, SiO, and SiS, originate close to the stellar photosphere and are released into the expanding outflow. These species are often classified as parent molecules \citep[see][]{Woods_et_al_2003, Agundez_et_al_2020}. In contrast, daughter species such as CN, HNC, and C\textsubscript{4}H \citep{Agundez_et_al_2017, Agundez_et_al_2020} form further out in the envelope, as a result of {photodissociation and subsequent chemical reactions}. Across the CSE, molecular abundances are modified by the complex interplay of processes such as gas-phase chemistry, dust-gas interactions, and photodissociation \citep[e.g.][]{Van_de_Sande_et_al_2023}. High-resolution observations and detailed modelling are required to trace the changing physical and chemical structure of these CSEs.

This work is the third in a series of papers that utilise the high angular resolution and sensitivity of the Atacama Large Millimeter/submillimeter Array (ALMA), to investigate the molecular chemistry in C-type AGB CSEs, extending beyond the well-studied, nearby carbon star IRC$+$10$\,$216 \citep[see e.g.][and references therein]{Agundez_et_al_2012, Agundez_et_al_2017, Cernicharo_et_al_2000, Pardo_et_al_2022feb, Siebert_et_al_2022, Velilla-Prieto_et_al_2015, Tuo_et_al_2024}, to observationally evaluate its often attributed status as an archetypal carbon star. \citet[][hereafter \citetalias{Unnikrishnan_et_al_2024}]{Unnikrishnan_et_al_2024} present spatially resolved, unbiased ALMA spectral surveys of three carbon stars, revealing similar chemical diversity and morphological complexity as IRC$+$10$\,$216 \citep{Velilla-Prieto_et_al_2019, Agundez_et_al_2017, Patel_et_al_2011, Cernicharo_et_al_2000}. \citet[][hereafter \citetalias{Unnikrishnan_et_al_2025}]{Unnikrishnan_et_al_2025} investigated the circumstellar dust and gas properties of five carbon stars, including IRC$+$10$\,$216 itself, using radiative transfer (RT) modelling of both dust and CO line emission. They also retrieved robust circumstellar radial abundance profiles of CS through non-local thermodynamic equilibrium (non-LTE) RT modelling, constrained by the ALMA data from \citetalias{Unnikrishnan_et_al_2024} and complementary single-dish {(SD)} observations.

Unlike CS, which can easily be formed under thermodynamical equilibrium in C-rich envelopes due to the presence of large amounts of carbon \citep[e.g.][]{Agundez_et_al_2020, Danilovich_et_al_2018}, the formation of silicon-bearing refractory species such as SiO and SiS may depend on several factors, {including freeze-out onto the dust and the possible need for shock chemistry \citep[e.g.][]{Cherchneff_2012} to form them in large amounts}. Both SD studies \citep[e.g.][]{Woods_et_al_2003, Schoier_et_al_2007, Agundez_et_al_2012, Van_de_Sande_et_al_2018_R_Dor, Danilovich_et_al_2018, Massalkhi_et_al_2019, Massalkhi_et_al_2024} and standalone modelling of interferometric data \citep[e.g.][]{Schoier_et_al_2006, Agundez_et_al_2017, Velilla-Prieto_et_al_2019} have been used to constrain the circumstellar abundances of parent species such as CS, SiO, and SiS. Though a few studies combining both SD and interferometric observations have been done for S- and M-type stars \citep[e.g.][]{Brunner_et_al_2018, Danilovich_et_al_2019}, to our knowledge, \citetalias{Unnikrishnan_et_al_2025} presents the first such models of C-type CSEs. Such integrated modelling, simultaneously employing both excitation and morphological constraints, represents a necessary advancement in accurately determining molecular abundances in CSEs. For this study, we modelled the species SiO and SiS, both of which are characterised by centrally peaked brightness distributions, using the same approach that was employed to model the CS emission in \citetalias{Unnikrishnan_et_al_2025} (see Sect.~\ref{sec:RT_modelling}).

\section{Source sample and observations}
\label{sec:Sample_and_observations}

\subsection{The sources}
\label{subsec:the_sources}
Our carbon star sample is the same as in \citetalias{Unnikrishnan_et_al_2025}, consisting of five high mass-loss rate (MLR), late-AGB phase sources, IRAS 15194$-$5115 (II Lup), IRAS 15082$-$4808 (V358 Lup), IRAS 07454$-$7112 (AI Vol), AFGL 3068 (LL Peg), and IRC~+10\,216 (CW Leo). The basic details of the sources are listed in Table~\ref{tab:source_properties}. The physical properties of the sources, derived using dust and CO RT modelling, have been presented in \citetalias{Unnikrishnan_et_al_2025}.

\begin{table*}[t]
   \caption{Basic source properties.}
   \label{tab:source_properties}
   \centering
      \begin{tabular}{l c c c c c}
      \hline\hline & \\[-2ex]
       \makecell{Source} & \makecell{{$\dot{M}_{{\rm H}_2}$$^{(a)}$}\\{($M_\odot$ yr$^{-1}$)}} & \makecell{Distance$^{(b)}$\\(pc)} &  \makecell{$\varv_\mathrm{sys}$\\(\small LSRK, km s$^{-1}$)} & \makecell{$\varv_\mathrm{\infty}$\\(\small LSRK, km s$^{-1}$)} & \makecell{$\dot{M}_{{\rm H}_2}$/$\varv_\mathrm{\infty}$\\($M_\odot$ yr$^{-1}$ km$^{-1}$ s)}\\
      \hline & \\[-2ex]
      IRAS 15194$-$5115 & $2.2\times10^{-5}$ & 696  & $-$15.0 & 21.5 & $1.0\times10^{-6}$\\
      IRAS 15082$-$4808 & $2.2\times10^{-5}$ & 1050 & $-$3.3  & 19.5 & $1.1\times10^{-6}$\\
      IRAS 07454$-$7112 & $8.3\times10^{-6}$ & 583  & $-$38.7 & 13.0 & $6.4\times10^{-7}$\\
      AFGL 3068    & $4.2\times10^{-5}$ & 1220 & $-$30.0 & 14.0 & $3.0\times10^{-6}$\\
      IRC~+10\,216 & $1.5\times10^{-5}$ & 190  & $-$26.5 & 14.5 & $1.0\times10^{-6}$\\
      \hline
      \end{tabular}
      \tablefoot{$\dot{M}_{{\rm H}_2}$ is the H$_2$ mass-loss rate, and $\varv_\mathrm{sys}$ and $\varv_\mathrm{\infty}$ denote the systemic and expansion velocities, respectively. The full list of stellar and CSE properties that are inputs to the RT models is given in \citetalias{Unnikrishnan_et_al_2025} (see their Table~3). $^{(a)}$\citetalias{Unnikrishnan_et_al_2025}; $^{(b)}$\citet{Andriantsaralaza_et_al_2022}.}
\end{table*}

\subsection{Spectral line surveys}
\label{subsec:spectral_line_surveys}
Multi-aperture spectra were extracted from our ALMA band 3 ($85-116$\,GHz) spectral survey cubes of SiO, SiS, and their {major} isotopologues {($^{29}$SiO, $^{30}$SiO, $^{29}$SiS, $^{30}$SiS and Si$^{34}$S)}, following the same procedure as in \citetalias{Unnikrishnan_et_al_2025}. {The line cubes were convolved to progressively larger circular Gaussian beams, with full width at half maximum (FWHM) beginning at the major-axis size of the original synthesised beam and gradually increasing until the resulting line flux reached a plateau, implying that all detected emission has been recovered. Extracting spectra from these convolved cubes allows tracing the line fluxes across different spatial scales and map the radial variations in line intensities. The spectral line profiles extracted from these cubes represent those that would result from observations using an SD telescope with the corresponding Gaussian beam size. The beam sizes listed along with the ALMA spectra in Figs.~\ref{fig:15194-5115_SiO}, \ref{fig:15194-5115_SiS}, and \ref{fig:15194-5115_29SiO} $-$ \ref{fig:10216_Si34S} denote the FWHMs of the circular Gaussian beams of the cubes from which the corresponding spectra were extracted}.
{Comparing with archival {SD} observations \citep{Woods_et_al_2003}, we find no evidence of resolved out flux in the lines of these species in our ALMA observations}. We complement these data with SD SiO and SiS lines from our APEX surveys. For IRAS 15194$-$5115, we also use lines from our \textit{Herschel}/HIFI survey. For all details of our ALMA, APEX, and HIFI observations, we refer the reader to \citetalias{Unnikrishnan_et_al_2024} and \citetalias{Unnikrishnan_et_al_2025}. The lines used in this work are listed in Tables~\ref{tab:SiO_line_intensities} $-$ \ref{tab:Si34S_line_intensities}.

\subsection{Supplementary observations}
\label{subsec:literature_data}
 We used SiO and SiS lines obtained from the literature as additional constraints to our RT models (Tables~\ref{tab:SiO_line_intensities} $-$ \ref{tab:Si34S_line_intensities}). For IRAS 07454$-$7112 and AFGL 3068, we used Atacama Compact Array (ACA) observations of SiO $J = 5-4$, SiS $J = 12-11$, $^{29}$SiS $J = 13-12$, and $^{30}$SiS $J = 19-18$, from the DEATHSTAR project \citep{Ramstedt_et_al_2020, Andriantsaralaza_et_al_2021}. The $^{29}$SiO $J = 8-7$ line was also obtained towards IRAS 07454$-$7112 from the ACA data,  but was not detected for AFGL 3068. The synthesised beam size of the ACA observations was {comparable} to the size of the molecular emitting regions, {meaning that all emission is recovered, but} we could not extract any information about the radial emission distribution using multi-aperture spectra as we do with the ALMA 12m array data. Hence, for use in constraining RT models, we treat these lines as if they were single-dish observations.

For IRAS 15194$-$5115 and IRAS 07454$-$7112, we compared our APEX SiS line spectra to those from \citet{Danilovich_et_al_2018}, and found them to be identical within the calibration uncertainties. For IRAS 07454$-$7112, we include in this work the APEX SiS $J = 16-15$ and $19-18$ lines from \citet{Danilovich_et_al_2018} which were not covered in our APEX survey.

For IRC+10216, we use interferometric line cubes that combined ALMA observations with IRAM 30m on-the-fly (OTF) maps, for the $J = 2-1$ lines of SiO, $^{29}$SiO, and $^{30}$SiO, and the $J = 5-4$ and $6-5$ lines of SiS, $^{29}$SiS, $^{30}$SiS, and Si$^{34}$S, from \citet{Velilla-Prieto_et_al_2019}. As this nearby ($\sim$190 pc, see Table~\ref{tab:source_properties}) source is very extended on the sky, the OTF observations help recover all the emission even at the most extended scales, that would be resolved out in the standalone ALMA observations, which have a maximum recoverable scale (MRS) of $\sim$23-32$\arcsec$ \citep{Velilla-Prieto_et_al_2019}. We also obtained a set of SD spectra observed using the IRAM 30m telescope for both SiO and SiS towards IRC+10216, from \citet{Agundez_et_al_2012}. These lines included SiO lines from $J = 2-1$ to $8-7$, and SiS lines $J = 5-4,~6-5$ and $J = 8-7$ to $19-18$. We also used the SiO $J = 1-0$ and the SiS $J = 2-1$ lines for IRC+10216, observed using the Yebes 40m telescope, from \citet{Massalkhi_et_al_2024}. In addition to these, we obtained archival SiS $J = 5-4$ ALMA observations towards IRC+10216 (project 2015.1.01271.S, PI: D. Keller), which contained a total power map, along with ALMA 12m and ACA observations. We combined these datasets using the same process as employed in \citetalias{Unnikrishnan_et_al_2024}, and extracted multi-aperture spectra from the combined cube.

\section{Radiative transfer modelling}
\label{sec:RT_modelling}

\subsection{Molecular data}
\label{subsec:Molecular_data}
{We take into account both radiative and collisional excitation in our RT modelling of both SiO and SiS. For the radiative part,} in our $^{28}$SiO, $^{29}$SiO, $^{30}$SiO models, we include energy levels $J$ = 0 $-$ 40, from both the $v = 0$ and $v = 1$ vibrational levels, leading to a total of 82 levels and 160 radiative transitions. Energies and radiative transition data were sourced from the Jet Propulsion Laboratory (JPL) spectroscopic database\footnote{\url{https://spec.jpl.nasa.gov}} \citep{Pickett_et_al_1998}. The molecular data for $^{28}$SiS and its {studied} isotopologues ($^{29}$SiS, $^{30}$SiS and Si$^{34}$S) are from the Cologne Database for Molecular Spectroscopy\footnote{\url{https://cdms.astro.uni- koeln.de}} \citep[CDMS,][]{Muller_et_al_2005, Endres_et_al_2016} catalogue, and are the same as used by \citet{Danilovich_et_al_2019}. For $^{28}$SiS, we include energy levels from $J$ = 0 to 99 from each of the vibrational levels $v = 0,~1,$ and 2, amounting to a total of 300 levels and 891 radiative transitions, including ro-vibrational transitions. For the other SiS isotopologues {studied in this work}, we included energy levels from $J$ = 0 to 40 in the vibrational levels $v = 0$ and $v = 1$, totalling 82 levels, with 160 radiative transitions. 

{For collisional excitation,} the $v=0$ SiO-H$_2$ and SiS-H$_2$ collisional rates, scaled and extrapolated from the SiO-H$_2$ rates by \citet{Dayou_and_Balanca_2006}, covering levels up to $J  = 40$ and giving a total of 820 collisional rate coefficients, were used for all {studied} isotopologues. Our RT code can only take into account one collision partner and hence we have not taken into consideration collisions with other partners, such as He. However, we do not expect this to significantly affect the modelling results, as the number density of He is expected to be only $\sim$17\% of that of H$_2$ {\citep[assuming solar number densities, e.g.][]{Asplund_et_al_2009}, and given the significant contribution of radiative excitation to the studied lines (see Sect.~\ref{subsec:ir_pumping})}.

\subsection{Modelling procedure}
\label{subsec:Modelling_procedure}
We adopt a spherically symmetric, smooth CSE model, with an isotropic, constant MLR as in \citetalias{Unnikrishnan_et_al_2025}. The stellar, dust, and gas input parameters for the RT models are taken from the SED fitting and CO modelling results presented in \citetalias{Unnikrishnan_et_al_2025} (see their Table~3). {The MLRs (Table~\ref{tab:source_properties}) used as inputs to the RT code are those of molecular hydrogen, H$_2$-MLR, and they are derived in \citetalias{Unnikrishnan_et_al_2025}. The circumstellar number densities used in the RT analysis are consequently H$_2$ densities. To a first approximation, assuming solar mass fractions, the total gas MLR, considering also the contribution of He, can be roughly estimated from the H$_2$-MLR by multiplying it by a factor of $\sim$1.4 (the reciprocal of the solar Hydrogen mass fraction). However, we have not made such a correction in this work as the scaling factor falls well within the estimated uncertainties on the derived MLRs \citepalias[factor of $\sim$2$-$3, see][]{Unnikrishnan_et_al_2025}}.

We use the latest version of the 1D accelerated lambda iteration (ALI) non-LTE spectral line RT code \citep{Maercker_et_al_2008, Danilovich_et_al_2018, Danilovich_et_al_2019} to model the line emission in this work. {All fractional abundances used in this work are defined as relative abundances with respect to molecular hydrogen (H$_2$), as in \citetalias{Unnikrishnan_et_al_2025}}. We use Gaussian abundance profiles, given by
\begin{equation}
f(r) = f_0 \exp\left(-\left(\frac{r}{R_e}\right)^2\right),
\label{eq:gaussian_abundance_profile}
\end{equation}
with the peak abundance ($f_0$) and the $e$-folding radius ($R_\mathrm{e}$) as free parameters in the modelling. To find the best-fit models, we employ $\chi^2$ minimisation {between the modelled and observed integrated line intensities}, taking into account the spatial constraints from the multi-aperture ALMA spectra, and covering a large range of excitation conditions traced by multiple SD lines. We refer to \citetalias{Unnikrishnan_et_al_2025} for a detailed description of the overall modelling technique and the {$\chi^2$ minimisation} methods adopted. In this work, we model all the {above} isotopologues with both $f_0$ and $R_\mathrm{e}$ as free parameters, and use the intersection of the respective $1\sigma$ $\chi^2$ contours of all {modelled} isotopologues per molecule, to determine the overall range of plausible models for each species.

\begin{figure*}[t]
    \centering
    \begin{subfigure}{0.375\textwidth}
        \includegraphics[width=\textwidth]{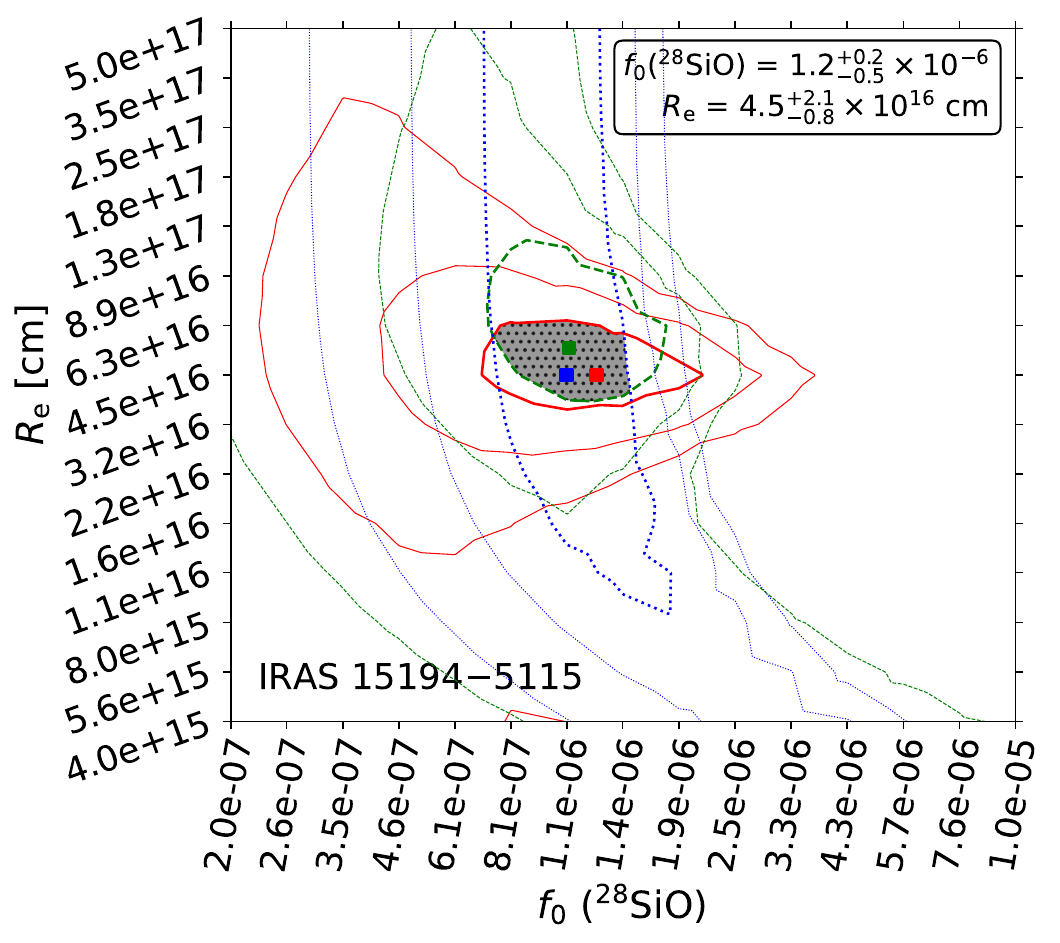}
        \caption{ IRAS 15194$-$5115}
        \label{subfig:SiO_chi_sq_map_IRAS_15194-5115}
    \end{subfigure}
    \hspace{0.01\textwidth}
    \begin{subfigure}{0.375\textwidth}
        \includegraphics[width=\textwidth]{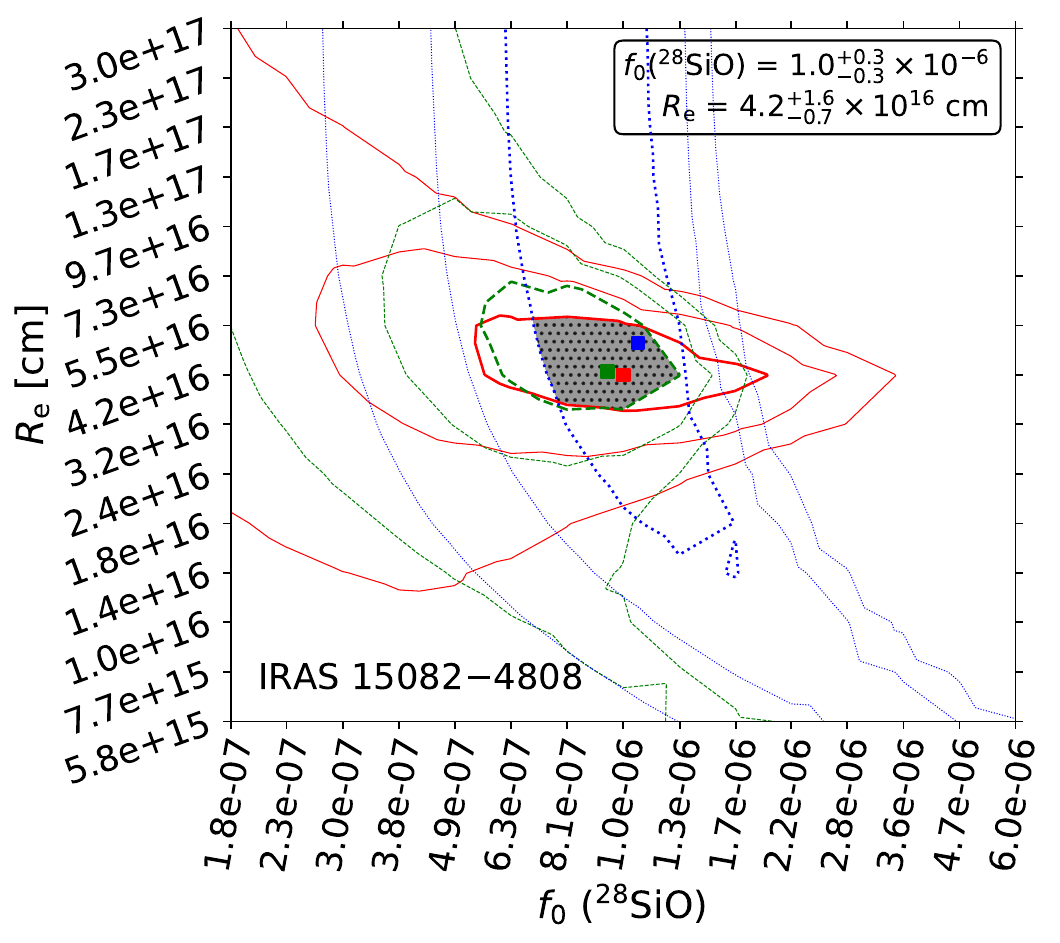}
        \caption{IRAS 15082$-$4808}
        \label{subfig:SiO_chi_sq_map_IRAS_15082-4808}
    \end{subfigure}

    \vspace{0.2em}

    \begin{subfigure}{0.375\textwidth}
        \includegraphics[width=\textwidth]{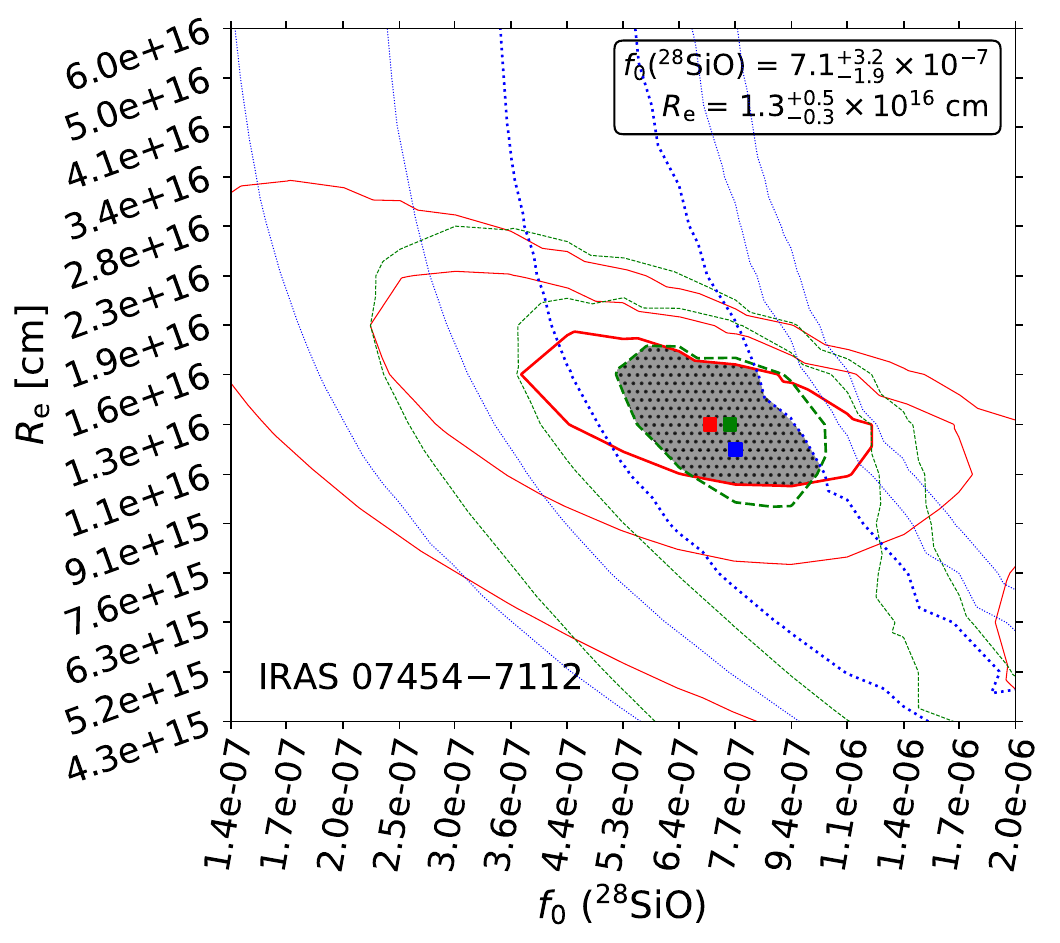}
        \caption{IRAS 07454$-$7112}
        \label{subfig:SiO_chi_sq_map_IRAS_07454-7112}
    \end{subfigure}
    \hspace{0.01\textwidth}
    \begin{subfigure}{0.375\textwidth}
        \includegraphics[width=\textwidth]{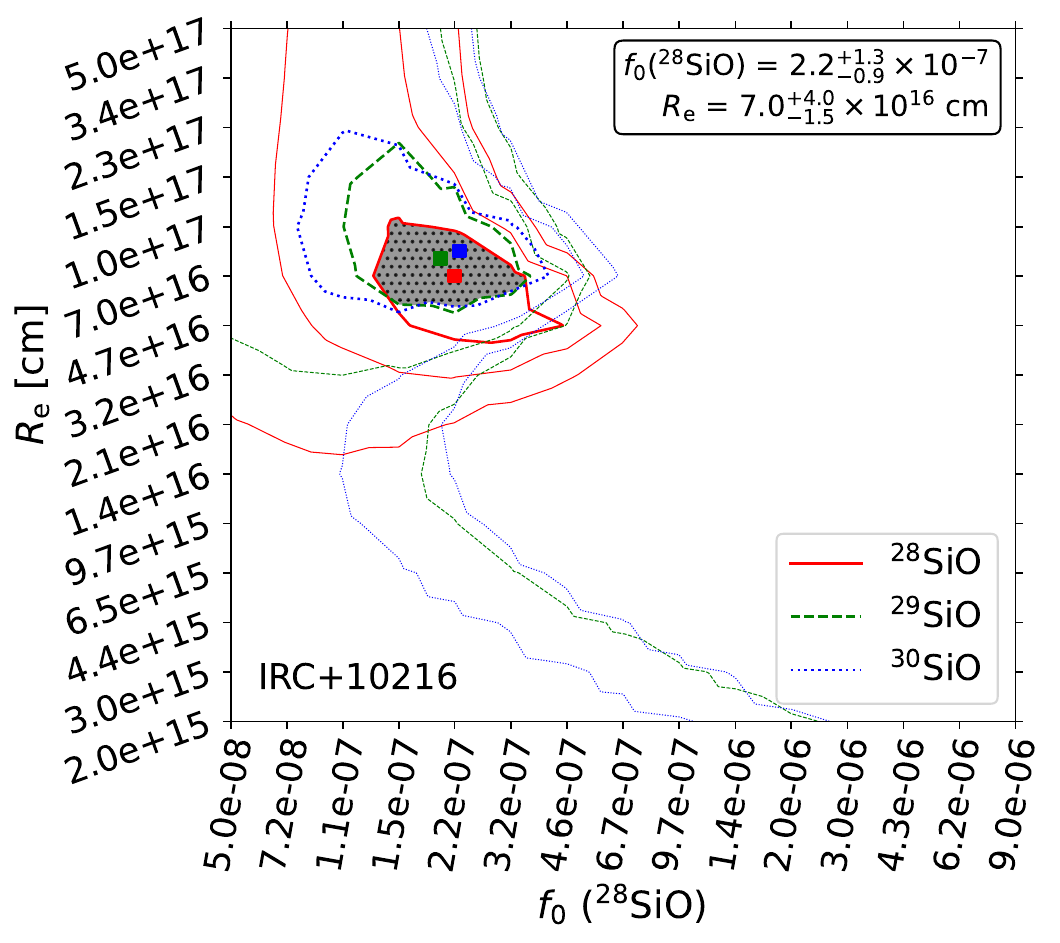}
        \caption{IRC~+10\,216}
        \label{subfig:SiO_chi_sq_map_IRC+10216}
    \end{subfigure}

    \caption{Grid results for SiO RT models for four of the sample stars. The 1, 2, and 3$\sigma$ contours in the $\chi^2$ space are shown for $^{28}$SiO (red), $^{29}$SiO (green), and $^{30}$SiO (blue). The thick lines mark the 1$\sigma$ contours, while the thin lines indicate the 2$\sigma$ and 3$\sigma$ contours. The hatched region represents the parameter space where the 1$\sigma$ contours of the three isotopologues overlap. The squares mark the `best-fit' models for the three isotopologues. The $x$-axis ticks show $^{28}$SiO abundances. The $^{29}$SiO and $^{30}$SiO abundances, at each point of the grid, are given by the corresponding $^{28}$SiO abundance divided by 20 and 30, respectively (see Sect.~\ref{subsec:Modelling_procedure}).}
    \label{fig:SiO_chi_sq_maps}
\end{figure*}

\begin{figure*}[t]
    \centering
    \includegraphics[width=0.975\linewidth]{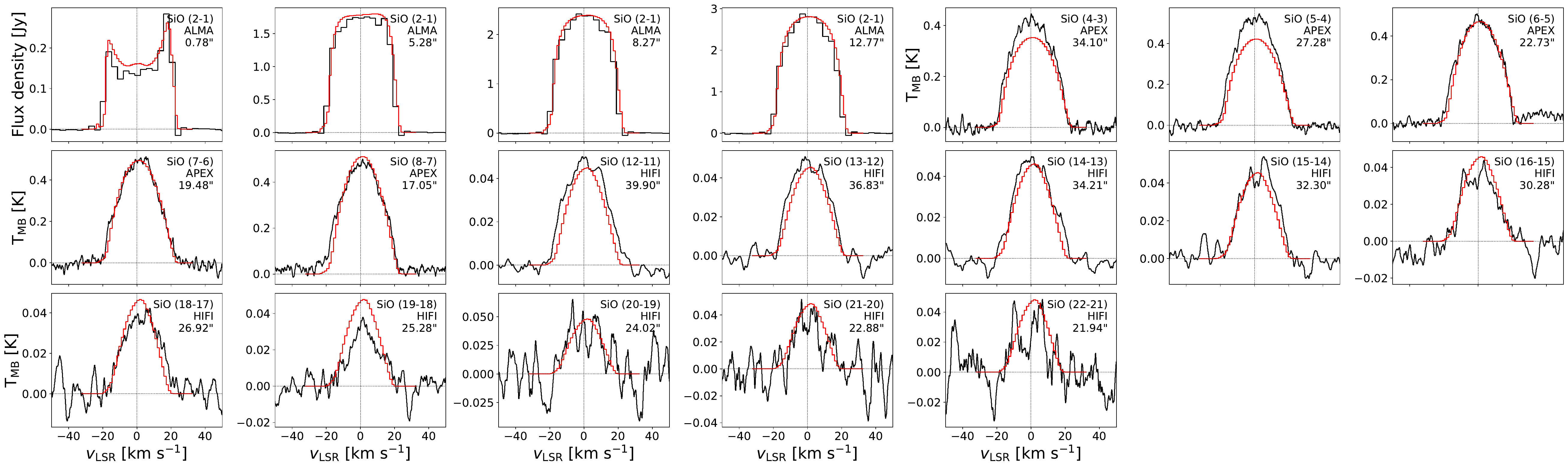}
    \caption{Observed (black) and modelled (red) SiO line profiles for IRAS 15194-5115. The transition quantum numbers, telescope used, and the beam size ({FWHM of the corresponding convolved Gaussian beam (see Sect.~\ref{subsec:spectral_line_surveys})} for the ALMA lines{; half power beam width (HPBW)} for the SD lines) of the observations are listed in the top right corner of each panel.}
    \label{fig:15194-5115_SiO}
\end{figure*}

The $f_0 - R_\mathrm{e}$ grids used for the five sources can be found in Figs.~\ref{fig:SiO_chi_sq_maps}, \ref{fig:SiS_chi_sq_maps}, and \ref{fig:AFGL_3068_chi_sq_maps}. The grids were chosen to encompass the expected ranges of SiO and SiS abundances and radial extents in carbon stars from previous SD estimates \citep[e.g.][]{Woods_et_al_2003, Agundez_et_al_2012, Danilovich_et_al_2018, Massalkhi_et_al_2019, Massalkhi_et_al_2024} and chemical modelling studies \citep[e.g.][]{Maes_et_al_2023, Van_de_Sande_et_al_submitted}. We note that though the upper limits of the $R_\mathrm{e}$ space sampled sometimes become comparable to the corresponding CO photodissociation radii, the determined best-fit values of $R_\mathrm{e}$ for all sources fall well within the corresponding CO radial extent, as expected \citepalias[see Table~\ref{tab:RT_modelling_results}, and][]{Unnikrishnan_et_al_2025}. The grids for the less-abundant isotopologues were designed to have the same $R_\mathrm{e}$ sampling as those of the respective main isotopologue grid. These species are photodissociated by radiation across the broadband continuum, and hence do not exhibit differential self-shielding between their isotopologues. The $f_0$ grids used for the less-abundant isotopologues were obtained by dividing those of the main isotopologues by the respective solar isotopic ratios, i.e. 20 for $^{28}$Si/$^{29}$Si, 30 for $^{28}$Si/$^{30}$Si, and 22 for $^{32}$S/$^{34}$S \citep{Asplund_et_al_2009}. The grid coverage was manually refined where needed to ensure that the total parameter space explored was adequately spanned and sampled. 

\section{Results}
\label{sec:Results}

\subsection{SiO}
\label{subsec:sio_results}
Our models constrain the SiO abundance profiles well for all sources except AFGL 3068, for which no spatially resolved information is available. The $\chi^2$ contour maps for SiO and its {studied} isotopologues for IRAS 15194$-$5115, IRAS 15082$-$4808, IRAS 07454$-$7112, and IRC~+10\,216 are shown in Fig.~\ref{fig:SiO_chi_sq_maps}. The best-fit $^{28}$SiO abundance profiles for the five sources derived from RT modelling are shown in Fig.~\ref{fig:CM_RT_abundance_comparisons}, and the corresponding $f_0$ and $R_\mathrm{e}$ values are given in Table~\ref{tab:RT_modelling_results}. We note that for $^{30}$SiO, for all five sources, we only use the $J = 4 - 3$ and $J = 6 - 5$ lines, as the $J = 5 - 4$ line is blended with the Si$^{34}$S $J = 12 - 11$ line. The sections below present the results of the SiO modelling. 

\subsubsection{IRAS 15194$-$5115 and IRAS 07454$-$7112}
\label{subsubsec:sio_results_15194_07454}
Models that reproduce the observed $^{28}$SiO, $^{29}$SiO, and $^{30}$SiO emission very well, across both the multi-aperture ALMA spectra and the SD line profiles, were obtained for both IRAS 15194$-$5115 and IRAS 07454$-$7112 (Figs.~\ref{fig:15194-5115_SiO}, \ref{fig:15194-5115_29SiO}, \ref{fig:IRAS-15194-5115-30SiO}, \ref{fig:IRAS-07454-7112-SiO}, \ref{fig:IRAS-07454-7112-29SiO}, \ref{fig:IRAS-07454-7112-30SiO}), yielding well-constrained abundance profiles (Fig.~\ref{fig:CM_RT_abundance_comparisons}). For $^{28}$SiO and $^{29}$SiO, we have both ALMA and SD lines, whereas for $^{30}$SiO, only SD lines are available for these stars. The $^{29}$SiO $J = 2 - 1$ line displays a sharp peak around $\sim$3-5 km s$^{-1}$ offset from the systemic velocity on the redshifted side for both sources, possibly {caused by substructure within the CSE} (Figs.~\ref{fig:15194-5115_29SiO} and \ref{fig:IRAS-07454-7112-29SiO}). {We note that it is unclear whether this could be a maser feature. Maser emission has been detected in the $^{29}$SiO $v=0, J=2-1$ line for several M-type stars \citep[e.g.][]{Deguchi_et_al_1983, Nguyen-Quang-Rieu_et_al_1988}, but not so far for carbon stars}.

\subsubsection{AFGL 3068}
\label{subsubsec:sio_results_3068}
For AFGL 3068, we only have 4 SD SiO lines available (Fig.~\ref{fig:AFGL-3068-SiO}), and no spatially resolved observations, leading to neither $f_0$ nor $R_\mathrm{e}$ being properly constrained (Fig.~\ref{subfig:AFGL_3068_chi_sq_map_SiO}). We also did not detect any $^{29}$SiO and $^{30}$SiO lines for this source. We produced modelled line profiles for the non-detected $^{29}$SiO and $^{30}$SiO $J = 4 - 3,~5 - 4$, and $6 - 5$ lines, which fall within the frequency range of our APEX observations, for a single abundance profile obtained by scaling down the derived $^{28}$SiO peak abundance (Table~\ref{tab:RT_modelling_results}) by the respective solar isotopic ratios (Table~\ref{tab:isotopic_ratios}), keeping the $e$-folding radius the same. These modelled line intensities were well within the noise of the observed spectra.

\subsubsection{IRC~$+$10\,216 and IRAS 15082$-$4808}
\label{subsubsec:sio_results_10216}
The SiO $J = 2 - 1$ line for IRC~+10\,216 has an asymmetric line profile, with a bump at the redshifted edge, possibly arising from sub-structure within the envelope (see Fig.~\ref{fig:IRC+10216-SiO}). For IRC~+10\,216, and also for IRAS 15082$-$4808 though to a much smaller extent, we find that the ALMA SiO $J = 2 - 1$ and the higher-$J$ SiO SD line profiles cannot be fit very well simultaneously. The models that fit the multi-aperture ALMA spectra well always underestimate the higher-$J$ SD lines (see Figs.~\ref{fig:IRAS-15082-4808-SiO}, \ref{fig:IRC+10216-SiO}), and conversely, the models that fit the higher-$J$ lines tend to over-predict the ALMA $J = 2 - 1$ line intensities. For IRC~+10\,216, this issue can also be seen in the models of the less-abundant SiO isotopologues (see Figs.~\ref{fig:IRC+10216-29SiO}, \ref{fig:IRC+10216-30SiO}). For IRAS 15082$-$4808, this is seen only for the main isotopologue (Fig.~\ref{fig:IRAS-15082-4808-SiO}), and not for $^{29}$SiO and $^{30}$SiO which are fit very well by our models (Figs.~\ref{fig:IRAS-15082-4808-29SiO}, \ref{fig:IRAS-15082-4808-30SiO}). 

For IRC~+10\,216, our best model for the SiO $J = 2 - 1$ ALMA spectra (see Fig.~\ref{fig:IRC+10216-SiO}) underestimates the APEX SiO $J = 4 - 3, 5 - 4, 6 - 5$ lines by $\sim$40\%. While this model fits the IRAM 30m SD spectrum of the SiO $J = 2 - 1$ line very well, consistent with its good fit of the $J = 2 - 1$ ALMA line, it underestimates the other IRAM 30m lines by $\sim$20-30\%. While this is still within a reasonable range, given the calibration uncertainties of the IRAM 30m ($\lesssim$30\%, see \citetalias{Unnikrishnan_et_al_2025}), it does indicate the possibility of a systematic offset. We note that for IRC~+10\,216, since the data includes zero-spacing information from the IRAM 30m OTF maps combined with the ALMA observations (see Sect.~\ref{subsec:literature_data}), we do not expect any significant resolved out flux in the emission maps despite the large extent of the emission on the sky. We tried several approaches to address this mismatch in the line fits, as described in Sect.~\ref{subsubsec:sio_alma_sd_mismatch_addressing}.

\subsubsection{Addressing the ALMA - SD fit mismatch}
\label{subsubsec:sio_alma_sd_mismatch_addressing}
Since our ALMA and SD observations were obtained at different times \citepalias[see][]{Unnikrishnan_et_al_2024, Unnikrishnan_et_al_2025}, we checked if SiO lines can be highly variable in carbon star envelopes, and found that the intensity variations are very small for the SiO lines used in this work \citet{Cernicharo_et_al_2014}. So, we cannot attribute the mismatch to variability alone. Also, even if the line intensities are strongly correlated with the stellar light curve, the time delay in receiving the stellar light can be different for different $J$ lines, especially if their excitation regions peak at different radii (see e.g. Fig.~\ref{fig:sio_excitation_checks}), an effect that cannot be incorporated into our RT models, which are static. We also tried varying the input stellar luminosity \citepalias[originally 12000 $L_\odot$ for IRC~+10\,216, see][]{Unnikrishnan_et_al_2025} within the range 5000-16000 $L_\odot$, but found that this did not affect the modelled line profiles. However, we note that the possible influence of stellar atmospheric bands in the excitation of these lines has not been considered here, as our models describe the central star using only its luminosity and temperature. Using a detailed stellar atmosphere model instead of a blackbody is currently not possible using our RT code. 

Emission from the dust also contributes significantly to the circumstellar IR radiation field. We find that changing the dust optical depth affects all line profiles, and not just the relatively higher-$J$ ones. We tested changing the input dust opacities for amorphous carbon grains from those adopted from \citet{Suh_2000} \citepalias[see][]{Unnikrishnan_et_al_2025} to those of \citet{Preibisch_et_al_1993}, and found that this did not change the modelled line intensities. We note, however, that these two sets of opacities are not widely different. Major changes to the dust opacity, such as turning off the dust radiation field altogether, which we also tested, affect all modelled lines, but do not show significant differential effects between the various transitions (see also Sect.~\ref{subsec:ir_pumping}).

\begin{figure*}[t]
    \centering
    \begin{subfigure}{0.375\textwidth}
        \includegraphics[width=\textwidth]{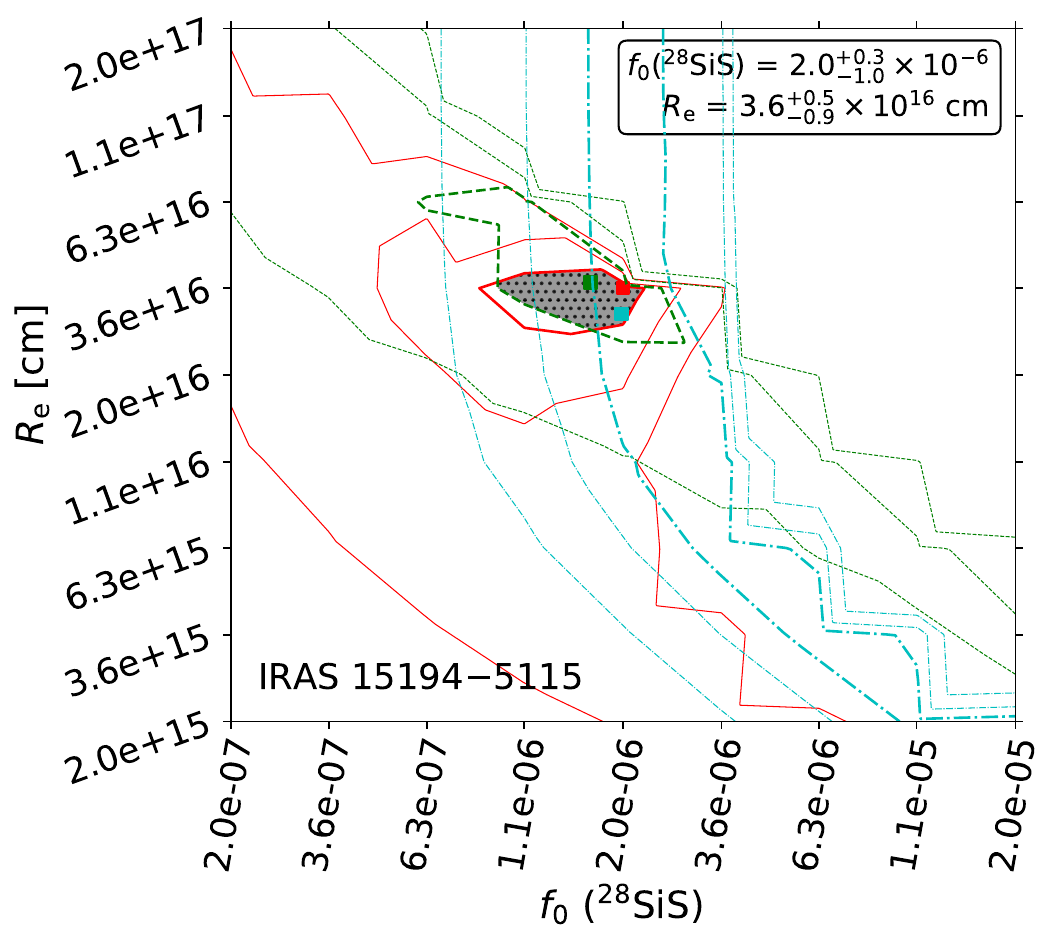}
        \caption{ IRAS 15194$-$5115}
        \label{subfig:SiS_chi_sq_map_15194}
    \end{subfigure}
    \hspace{0.01\textwidth}
    \begin{subfigure}{0.375\textwidth}
        \includegraphics[width=\textwidth]{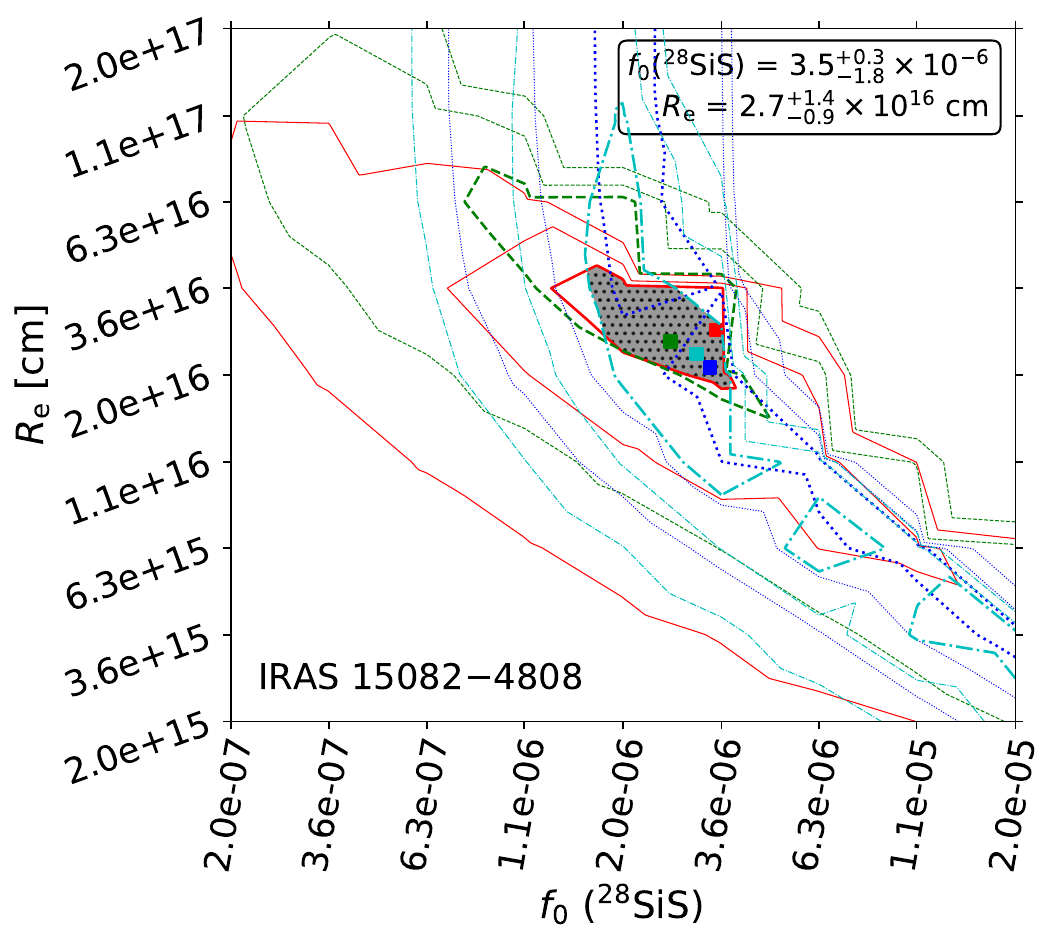}
        \caption{IRAS 15082$-$4808}
        \label{subfig:SiS_chi_sq_map_15082}
    \end{subfigure}

    \vspace{0.2em}

    \begin{subfigure}{0.375\textwidth}
        \includegraphics[width=\textwidth]{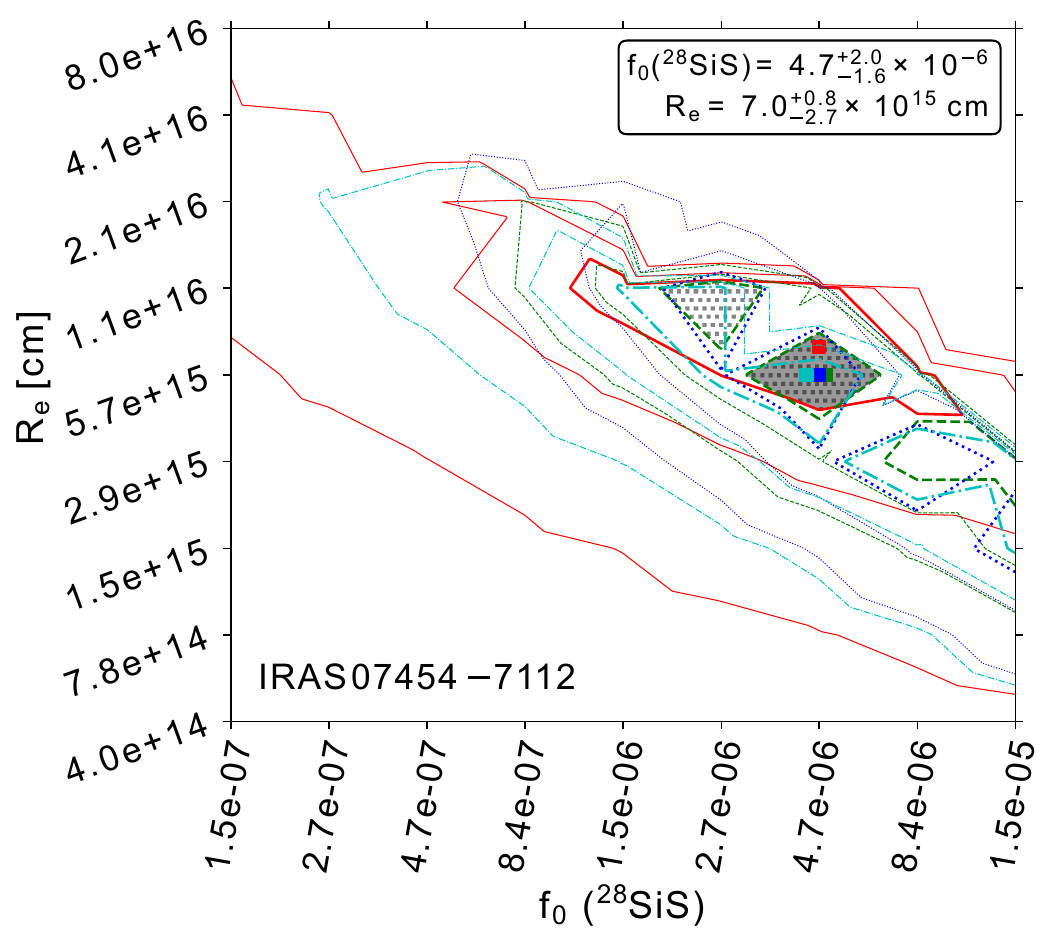}
        \caption{IRAS 07454$-$7112}
        \label{subfig:SiS_chi_sq_map_07454}
    \end{subfigure}
    \hspace{0.01\textwidth}
    \begin{subfigure}{0.375\textwidth}
        \includegraphics[width=\textwidth]{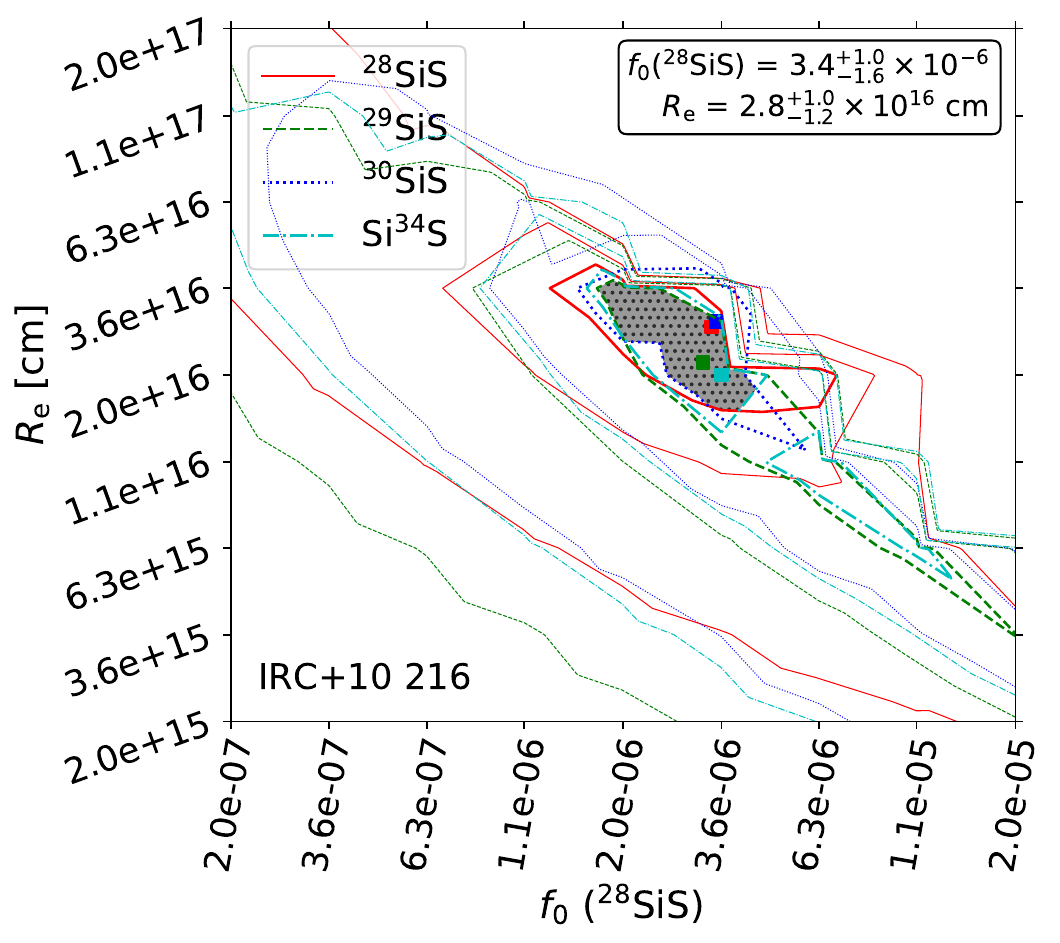}
        \caption{IRC~+10\,216}
        \label{subfig:SiS_chi_sq_map_10216}
    \end{subfigure}

    \caption{Grid results for SiS RT models for four stars of the sample. As in Fig.~\ref{fig:SiO_chi_sq_maps}, but for $^{28}$SiS (red), $^{29}$SiS (green), $^{30}$SiS (blue), and Si$^{34}$S (cyan). For IRAS 07454$-$7112, the contours shown are 2, 3, and 4 $\sigma$ (see Sect.~\ref{subsubsec:sis_results_07454}). See Sect.~\ref{subsec:sis_results} for descriptions of the hatched `best-fit' regions for the individual sources. The $x$-axis ticks show $^{28}$SiS abundances. The $^{29}$SiS, $^{30}$SiS, and Si$^{34}$S abundances at each point of the grid are given by its $^{28}$SiS abundance divided by 20, 30, and 22, respectively (see Sect.~\ref{subsec:Modelling_procedure}).}
    \label{fig:SiS_chi_sq_maps}
\end{figure*}

Further, in order to check if the addition of higher vibrational levels than those originally included (see Sect.~\ref{subsec:Molecular_data}) would affect the modelled line intensities by additional IR pumping through higher levels, we tested an SiO molecular description which included levels ranging from $v = 0 - 6$. The resulting modelled line profiles match those modelled using only the $v$ = 0,1 levels within the observational uncertainties.

We also attempted using modified abundance profiles, which had a central step function with a constant abundance $f_\mathrm{c}$ out to a radius $R_\mathrm{c}$, followed by the Gaussian profile at larger radii. Based on our IRC~+10\,216 model which fits the ALMA SiO $J = 2 - 1$ line well (see Table~\ref{tab:RT_modelling_results} and Fig.~\ref{fig:CM_RT_abundance_comparisons}), we ran a grid of models where the step function abundance in the inner part ($f_\mathrm{c}$) varies between 1.0$\times$10$^{-8}$ and 5.0$\times$10$^{-6}$, up to radial distances ($R_\mathrm{c}$) in the range 1.0$\times$10$^{15}$ - 2.0$\times$10$^{16}$ cm, subsequently followed by the original Gaussian abundance profile at larger radii. The usual logic behind such profiles is that the higher-$J$ lines, such as our SD lines here, are excited closer to the star than the comparatively lower-$J$ lines, such as the ALMA lines in this work, and hence any changes in the fractional abundance in the inner CSE could theoretically impact the intensity of the higher-$J$ lines without significantly affecting the low-$J$ lines \citep[see e.g.][]{Danilovich_et_al_2019}. However, in our case, this method was also unsuccessful in simultaneously fitting all the ALMA and SD line profiles, as there was no step-function - Gaussian combination which could successfully increase the intensity of the underestimated SD lines without simultaneously forcing the small-aperture ALMA lines to be significantly over-predicted. The implications of this are further discussed in Sect.~\ref{subsec:10216_SiO_fit_mismatch}. 

\begin{figure*}[t]
    \centering
    \includegraphics[width=0.975\linewidth]{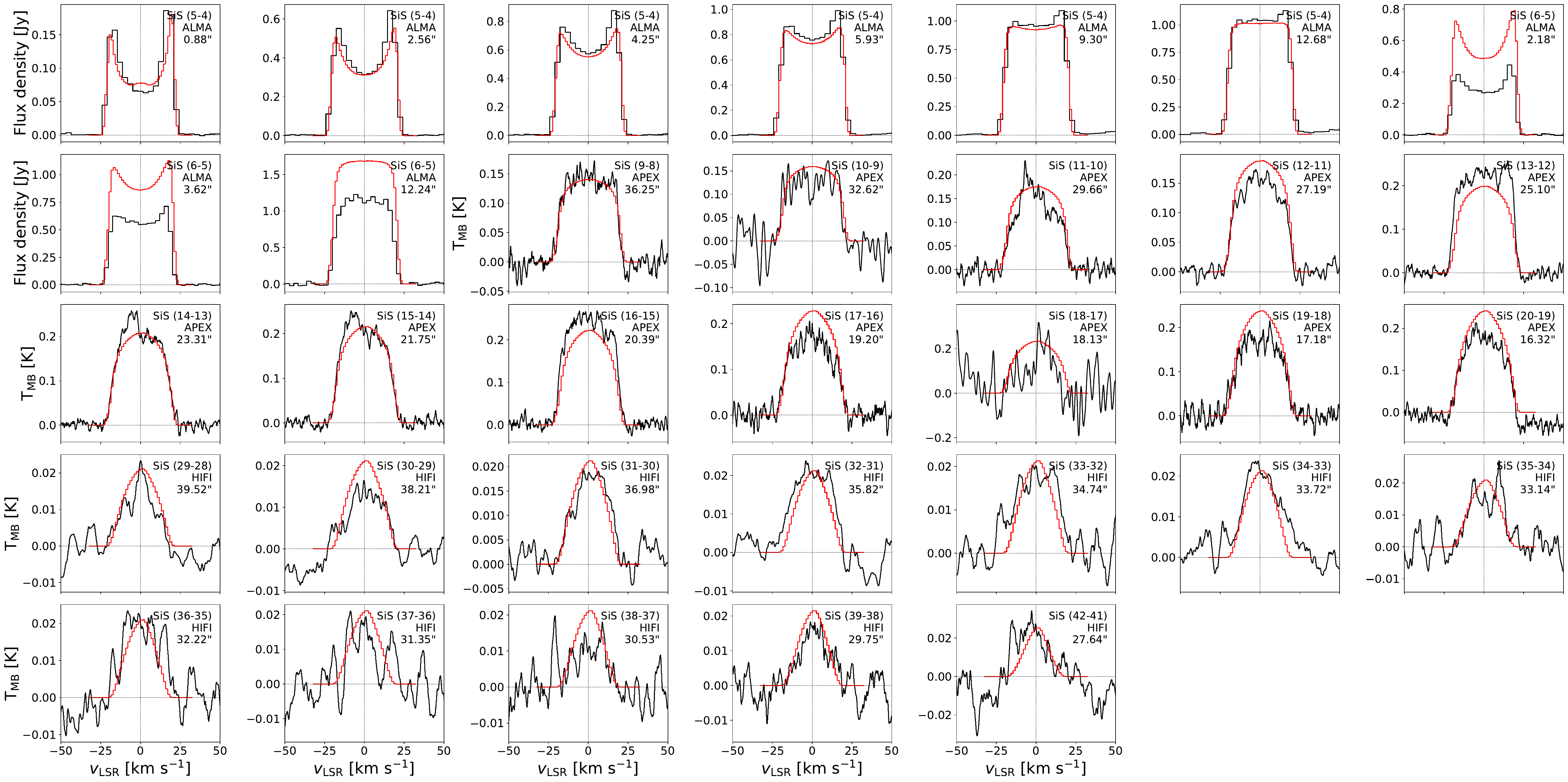}
    \caption{Observed (black) and modelled (red) SiS line profiles for IRAS 15194-5115. The transition quantum numbers, telescope used, and the beam size ({FWHM of the corresponding convolved Gaussian beam (see Sect.~\ref{subsec:spectral_line_surveys})} for the ALMA lines{;} HPBW for the SD lines) of the observations are listed in the top right corner of each panel.}
    \label{fig:15194-5115_SiS}
\end{figure*}

We also varied the exponent in our input abundance profile description (see Eq.~\ref{eq:gaussian_abundance_profile}), originally set to 2 for Gaussian profiles. We found that changing this value affects all lines, including the ALMA and SD ones, with the line intensities nominally increasing for increasing values, and hence cannot account for the mismatch seen between the modelled profiles of the different $J$ lines.

\begin{figure}[h]
    \centering
    \begin{subfigure}{0.3\textwidth}
        \includegraphics[width=\textwidth]{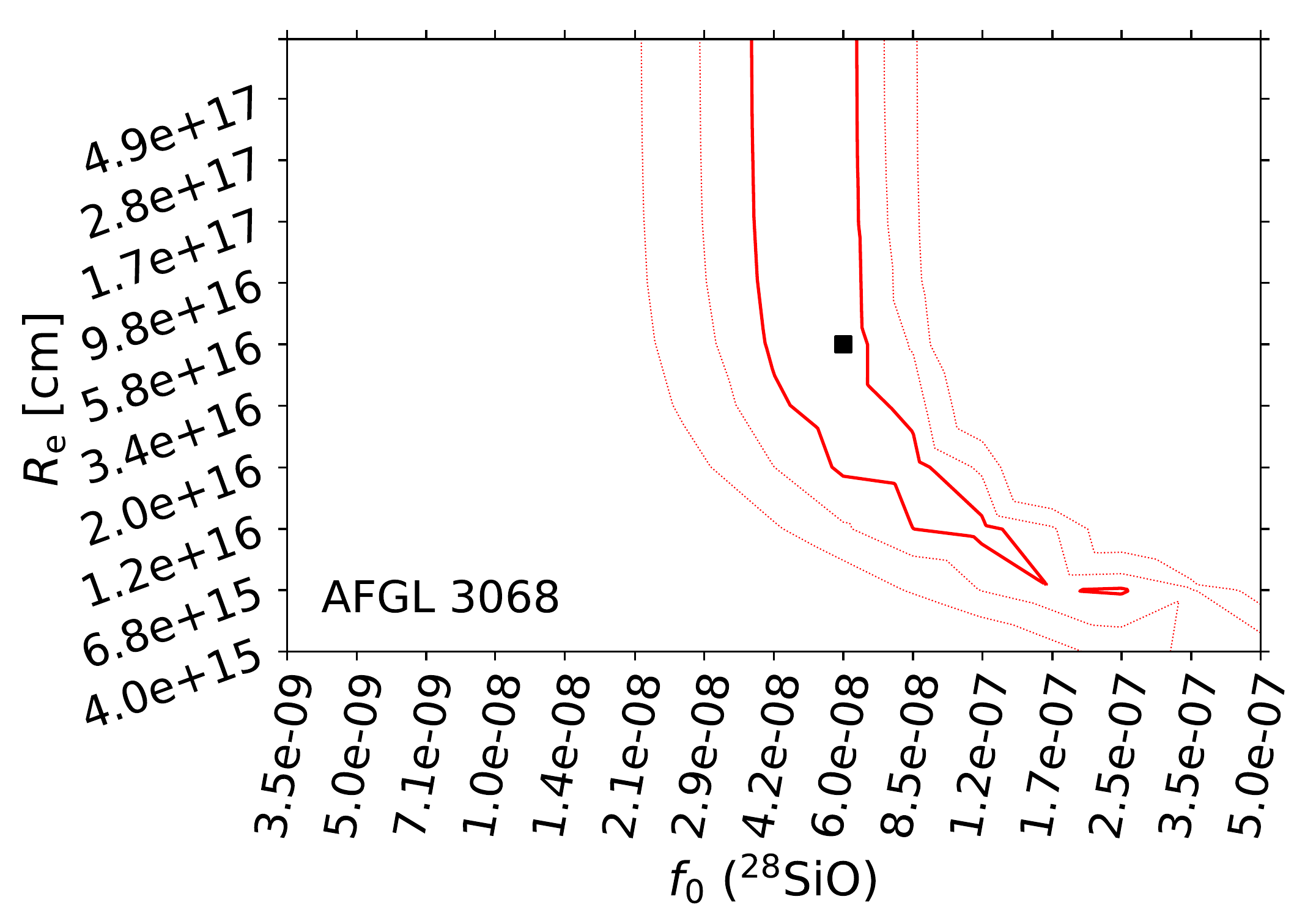}
        \caption{AFGL 3068 - $^{28}$SiO}
        \label{subfig:AFGL_3068_chi_sq_map_SiO}
    \end{subfigure}

    \begin{subfigure}{0.3\textwidth}
        \includegraphics[width=\textwidth]{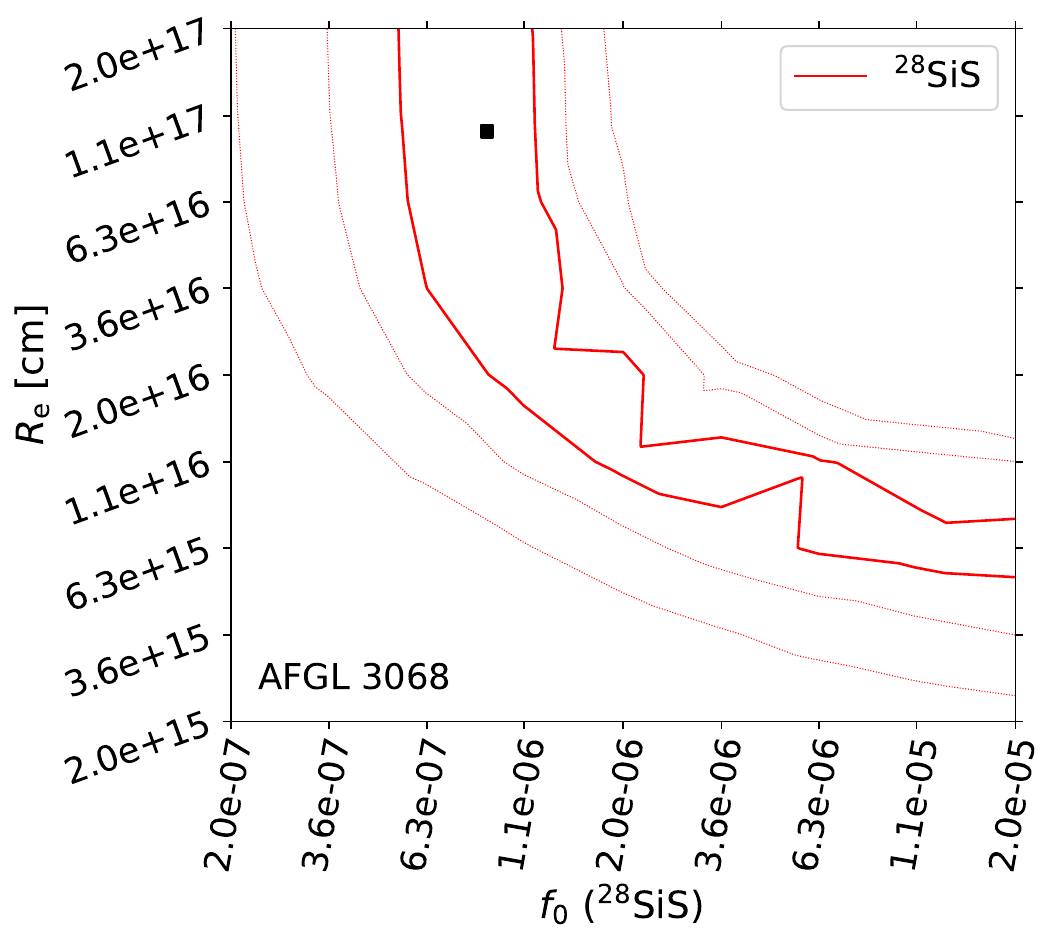}
        \caption{AFGL 3068 - $^{28}$SiS}
        \label{subfig:AFGL_3068_chi_sq_map_SiS}
    \end{subfigure}

    \caption{SiO and SiS grid results for AFGL 3068. The 1, 2, and 3$\sigma$ contours in the $\chi^2$ space are shown for (a) $^{28}$SiO and (b) $^{28}$SiS. The degeneracy and lack of strong constraints are due to the non-availability of spatially resolved data (see text). The black square marks the overall `best-fit' model.}
    \label{fig:AFGL_3068_chi_sq_maps}
\end{figure}

None of our IRC~+10\,216 SiO models that reasonably fit the ALMA lines manages to fit the Yebes SiO $J = 1 - 0$ line. Overall, for IRC~+10\,216, given the incompatibility between the ALMA and SD spectra, we choose to constrain our SiO and isotopologue models using only our ALMA spectra, instead of the SD data, as the ALMA data have much lower calibration uncertainties, and are also all observed at the same time.

\subsection{SiS}
\label{subsec:sis_results}
We were able to successfully constrain the SiS abundances for all sources except AFGL 3068. The $\chi^2$ contour maps for our grids of models for four of the sources, showing the overlaps in the best-fit regions of the different isotopologues, are given in Fig.~\ref{fig:SiS_chi_sq_maps}. For IRAS 15194$-$5115 and IRAS 07454$-$7112, we have only used the well-constrained $^{28}$SiS and $^{29}$SiS contours to determine the overlapping best-fit region of the different isotopologues (Figs.~\ref{subfig:SiS_chi_sq_map_15194}, \ref{subfig:SiS_chi_sq_map_07454}), as the contours of $^{30}$SiS and Si$^{34}$S are not very well constrained (Sections~\ref{subsubsec:sis_results_15194}, \ref{subsubsec:sis_results_07454}). For IRAS 15082$-$4808, the $^{28}$SiS, $^{29}$SiS, and Si$^{34}$S contours are used, leaving out the more uncertain $^{30}$SiS contours (Fig.~\ref{subfig:SiS_chi_sq_map_15082}, Sect.~\ref{subsubsec:sis_results_15082}). For IRC~+10\,216, where we have well-constrained abundances for all four isotopologues (Sect.~\ref{subsubsec:sis_results_10216}), we have used the 1$\sigma$ contours of all of them to determine the `best-fit' region in the $f_0$ - $R_\mathrm{e}$ parameter space (see Fig.~\ref{subfig:SiS_chi_sq_map_15082}). Fig.~\ref{fig:CM_RT_abundance_comparisons} shows the best-fit SiS abundance profiles for the sources in our sample. The corresponding $f_0$ and $R_\mathrm{e}$ values are listed in Table~\ref{tab:RT_modelling_results}.

The asymmetric horned profiles seen in the SiS $J = 5 - 4$ lines (Figs.~\ref{fig:15194-5115_SiS}, \ref{fig:IRAS-15082-4808-SiS}, \ref{fig:IRAS-07454-7112-SiS}, \ref{fig:IRC+10216-SiS}) for all four sources where we detect the line, are possibly due to weak maser emission \citep[e.g.]{Olofsson_et_al_1982, Sahai_et_al_1984, Fonfria_et_al_2018}. We note that for all sources, except AFGL 3068 where we do not have ALMA data, the ALMA spectra for the SiS $J = 5 - 4$ and $6 - 5$ lines cannot be simultaneously fit, as all models that fit the SiS $J = 5 - 4$ line significantly over-predict the SiS $J = 6 - 5$ line. The reasons for this are explored in more detail in Sect.~\ref{subsec:SiS_6_5_line}. We also note that while we are able to find models that fit the SiS $J = 5 - 4$ ALMA spectra and the available SD lines simultaneously for all sources, there are no models that fit the SiS $J = 6 - 5$ ALMA lines and the SD lines together. We therefore choose to constrain our models using the SiS $J = 5 - 4$ ALMA line and all available SD lines, and exclude the $J = 6 - 5$ lines from this analysis, for all sources.

{\renewcommand{\arraystretch}{1.5}%
\begin{table*}[t]
   \caption{Results of RT modelling for the main isotopologues of SiO and SiS.}
   \label{tab:RT_modelling_results}
   \centering
   \begin{tabular}{c@{\extracolsep{10pt}}cc@{\extracolsep{10pt}}cc}
      \hline\hline \\[-5.5ex]
      \makecell{\\\\Source} & 
      \multicolumn{2}{c}{SiO} &
      \multicolumn{2}{c}{SiS} \\[-2.0ex]
      \cline{2-3} \cline{4-5} \\[-4.0ex]
      & $f_0$ & $R_\text{e}$ [cm] & $f_0$ & $R_\text{e}$ [cm] \\[0.5ex]
      \hline \\[-3.5ex]
        IRAS 15194$-$5115  & $1.2^{+0.2}_{-0.5}\times10^{-6}$ & $4.5^{+2.1}_{-0.8}\times10^{16}$ & $2.0^{+0.3}_{-1.0}\times10^{-6}$ & $3.6^{+0.5}_{-0.9}\times10^{16}$ \\
        IRAS 15082$-$4808  & $1.0^{+0.3}_{-0.3}\times10^{-6}$ & $4.2^{+1.6}_{-0.7}\times10^{16}$ & $3.5^{+0.3}_{-1.8}\times10^{-6}$ & $2.7^{+1.4}_{-0.9}\times10^{16}$ \\
        IRAS 07454$-$7112  & $7.1^{+3.2}_{-1.9}\times10^{-7}$ & $1.3^{+0.5}_{-0.3}\times10^{16}$ & $4.7^{+2.0}_{-1.6}\times10^{-6}$ & $7.0^{+0.8}_{-2.7}\times10^{15}$ \\
        AFGL 3068     & $6.0^{+\uparrow}_{-\downarrow}\times10^{-8}$               & $5.8^{+\uparrow}_{-\downarrow}\times10^{16}$      & $9.0^{+\uparrow}_{-\downarrow}\times10^{-7}$         & $1.0^{+\uparrow}_{-\downarrow}\times10^{17}$ \\
        IRC~$+$10\,216 & $2.2^{+1.3}_{-0.9}\times10^{-7}$ & $7.0^{+4.0}_{-1.5}\times10^{16}$ & $3.4^{+1.0}_{-1.6}\times10^{-6}$ & $2.8^{+1.0}_{-1.2}\times10^{16}$ \\[0.5ex]
      \hline \\[-2ex]
   \end{tabular}
   \tablefoot{$f_0$: peak abundance, $R_\text{e}$: $e$-folding radius of the abundance profile. The sub/super-scripts show the uncertainties on the values, and $^{+\uparrow}_{-\downarrow}$ indicates that the uncertainties have not been constrained. {We note that the reported uncertainties do not take into account possible uncertainties in the input MLR and distance, which are taken as fixed parameters in the modelling}.}
\end{table*}}

We note that for Si$^{34}$S, we do not use the $J = 12 - 11$ line to constrain our models as it is blended with the $^{30}$SiO $J = 5 - 4$ line as mentioned in Sect.~\ref{subsec:sio_results}. Further, as done for SiO (see Sect.~\ref{subsubsec:sio_alma_sd_mismatch_addressing}), we tested an extended SiS dataset including the levels $v = 0 - 5$ along with the vibration-rotation transitions with $\Delta v = 1, 2$ for the $v=0, 1$, and 2 levels, and found that the resulting spectra matched with those modelled using only the $v = 0,~1, 2$ levels. The following sections describe the SiS modelling results for the different sources.

\subsubsection{IRAS 15194-5115}
\label{subsubsec:sis_results_15194}
For IRAS 15194$-$5115, we model the SiS isotopologues $^{28}$SiS, $^{29}$SiS, and Si$^{34}$S, but not $^{30}$SiS which is not detected towards this source. We find very good fits to the observed line profiles of the different lines for all three modelled isotopologues (see Figs.~\ref{fig:15194-5115_SiS}, \ref{fig:IRAS-15194-5115-29SiS}, \ref{fig:IRAS-15194-5115-Si34S}), though the uncertainties on the Si$^{34}$S abundances are not well-constrained (Fig.~\ref{subfig:SiS_chi_sq_map_15194}). For $^{28}$SiS and Si$^{34}$S, our best-fit models over-predict the $J = 6 - 5$ line (Figs.~\ref{fig:15194-5115_SiS}, \ref{fig:IRAS-15194-5115-Si34S}). We note that for $^{29}$SiS towards this source, we do not have any SD lines available, but only the $J = 5 - 4$ and $J = 6 - 5$ lines observed using ALMA. Our model overestimates the smallest aperture ALMA line profile for the $^{29}$SiS $J = 6 - 5$ line, whereas the larger apertures are well fit.

\subsubsection{IRAS 15082-4808}
\label{subsubsec:sis_results_15082}
Overall, we obtain reasonably well-constrained abundance profiles for this source. We do not have spatially resolved lines for $^{30}$SiS for this star, leading us to rely on only the two APEX spectra (Fig.~\ref{fig:IRAS-15082-4808-30SiS}). The $^{30}$SiS $\chi^2$ map has broader, less constrained 1$\sigma$ contours than the other isotopologues, indicating comparatively larger uncertainties (Fig.~\ref{subfig:SiS_chi_sq_map_15082}). The $^{29}$SiS $J = 5 - 4$ line presents a peculiar line profile with a central spike (Fig.~\ref{fig:IRAS-15082-4808-29SiS}), indicating either maser emission or possibly a blend with the HC$_7$N $v=0,~J = 79-78$ line, and is not fit very well by our models.

\subsubsection{IRAS 07454-7112}
\label{subsubsec:sis_results_07454}
We find models that fit the observed line emission well for all four SiS isotopologues {studied} for IRAS 07454$-$7112 (Figs.~\ref{fig:IRAS-07454-7112-SiS} - \ref{fig:IRAS-07454-7112-Si34S}). For this source, we use the overlapping region of the 2$\sigma$ contours of $^{28}$SiS and $^{29}$SiS to define the best-fit region (see Fig.~\ref{subfig:SiS_chi_sq_map_07454}), as the 1$\sigma$ contours covered only unrealistically narrow ranges. We find two distinct regions where the contours overlap, but determined by visual inspection of the modelled spectra that the overall line profile fits are better, especially for the SD lines, only for the filled-dotted region towards larger $f_0$ values, and not the dotted region (Fig.~\ref{subfig:SiS_chi_sq_map_07454}). Therefore, we use only the former area to define our best-fit region for SiS for this source. Our models fit well the ACA spectra for the $^{28}$SiS $J = 12 - 11$, and the $^{29}$SiS $J = 13 - 12$ lines, while underestimating the ACA $^{30}$SiS $J = 19 - 18$ line. We find that $^{28}$SiS requires a slightly larger $R_\mathrm{e}$ than the other isotopologues for this source (Fig.~\ref{subfig:SiS_chi_sq_map_07454}).

\subsubsection{IRC+10\,216}
\label{subsubsec:sis_results_10216}
For this source, we had two independent spatially resolved observations of the $^{28}$SiS $J = 5 - 4$ line (see Sect.~\ref{subsec:literature_data}). We were able to find models that simultaneously fit the multi-aperture spectra for these two lines, as well as a range of SD lines, including APEX observations ($J = 9 - 8$ to $14 - 13$) and the Yebes 40m $J = 2 - 1$ line (see Fig.~\ref{fig:IRC+10216-SiS}). However, this model overestimates several of the same SiS lines ($J = 9 - 8$ to $13 - 12$) and some other higher-$J$ lines ($J = 16 - 15$ to $19 - 18$) observed with the IRAM 30m telescope. However, the IRAM 30m SiS $J = 5 - 4$, $14 - 13$, and $15 - 14$ lines are fitted very well (Fig.~\ref{fig:IRC+10216-SiS}). 

For the isotopologues $^{29}$SiS, $^{30}$SiS, and Si$^{34}$S, for all of which we have both ALMA and SD lines available, our best models fit all available observations, except for some of the smaller apertures in the ALMA multi-aperture spectra for the $J = 6 - 5$ lines (see Figs.~\ref{fig:IRC+10216-29SiS}, \ref{fig:IRC+10216-30SiS}, \ref{fig:10216_Si34S}). The observed ALMA $J = 5 - 4$ and $6 - 5$ line profiles of $^{29}$SiS, $^{30}$SiS, and Si$^{34}$S for IRC~+10\,216 show a sharp peak around the systemic velocity in the small aperture spectra, which is not visible for larger apertures (see Figs.~\ref{fig:IRC+10216-29SiS}, \ref{fig:IRC+10216-30SiS}, \ref{fig:10216_Si34S}). This is probably formed due to gas velocity variations occurring very close to the star, or sub-structure within the circumstellar envelope. The Si$^{34}$S $J = 5 - 4$ line for this source also presents a horn at the redshifted end of the line profile (Fig.~\ref{fig:10216_Si34S}), similar to the one seen in the $^{28}$SiS $J = 5 - 4$ lines for all our sources (see e.g. Fig.~\ref{fig:15194-5115_SiS}).

\subsubsection{AFGL 3068}
\label{subsubsec:sis_results_3068}
For AFGL 3068, as in the case of SiO, we only have SD lines and no spatially resolved information available. Therefore, the SiS abundance profile of this source is also not well-constrained (see Fig.~\ref{subfig:AFGL_3068_chi_sq_map_SiS}). However, unlike in the case of SiO, we have SD lines detected for all four SiS isotopologues {used}, and are able to find models that fit all of them, with similar $R_\mathrm{e}$ values (Figs.~\ref{fig:AFGL-3068-SiS}, \ref{fig:AFGL-3068-29SiS}, \ref{fig:AFGL-3068-30SiS}, \ref{fig:AFGL-3068-Si34S}), and with the abundance ratios matching well with the expected isotopic ratios (see Table~\ref{tab:isotopic_ratios}). This gives more confidence to the derived abundance profiles, though we are not able to determine robust uncertainty ranges in this case, as we do for the other sources.

\begin{figure}[t]
    \centering
    \includegraphics[width=0.7\linewidth]{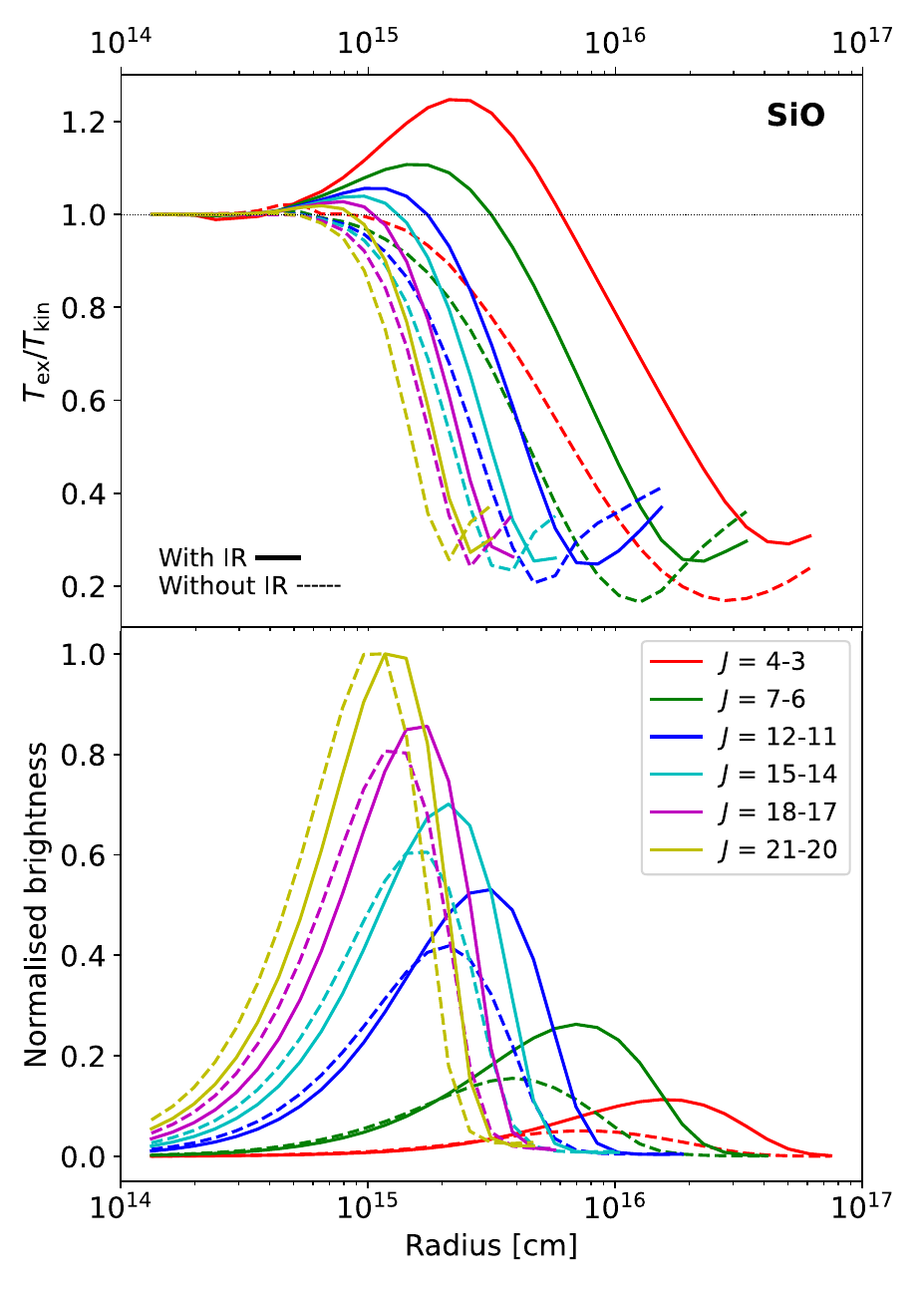}
    \caption{Ratio of the excitation temperature ($T_\mathrm{ex}$) and the kinetic temperature ($T_\mathrm{kin}$) for various low- and high-$J$ SiO transitions (top), and their corresponding line-emitting regions (bottom), as functions of radius across the CSE, from the best-fit model for IRAS 15194$-$5115. The solid lines are from a model that takes into account IR pumping, while the dashed lines are from a model that does not, but is otherwise identical. The line-emitting regions are normalised to the peak of the emitting region of the highest $J$ transition shown.}
    \label{fig:sio_excitation_checks}
\end{figure}

\begin{figure}[t]
    \centering
    \includegraphics[width=0.7\linewidth]{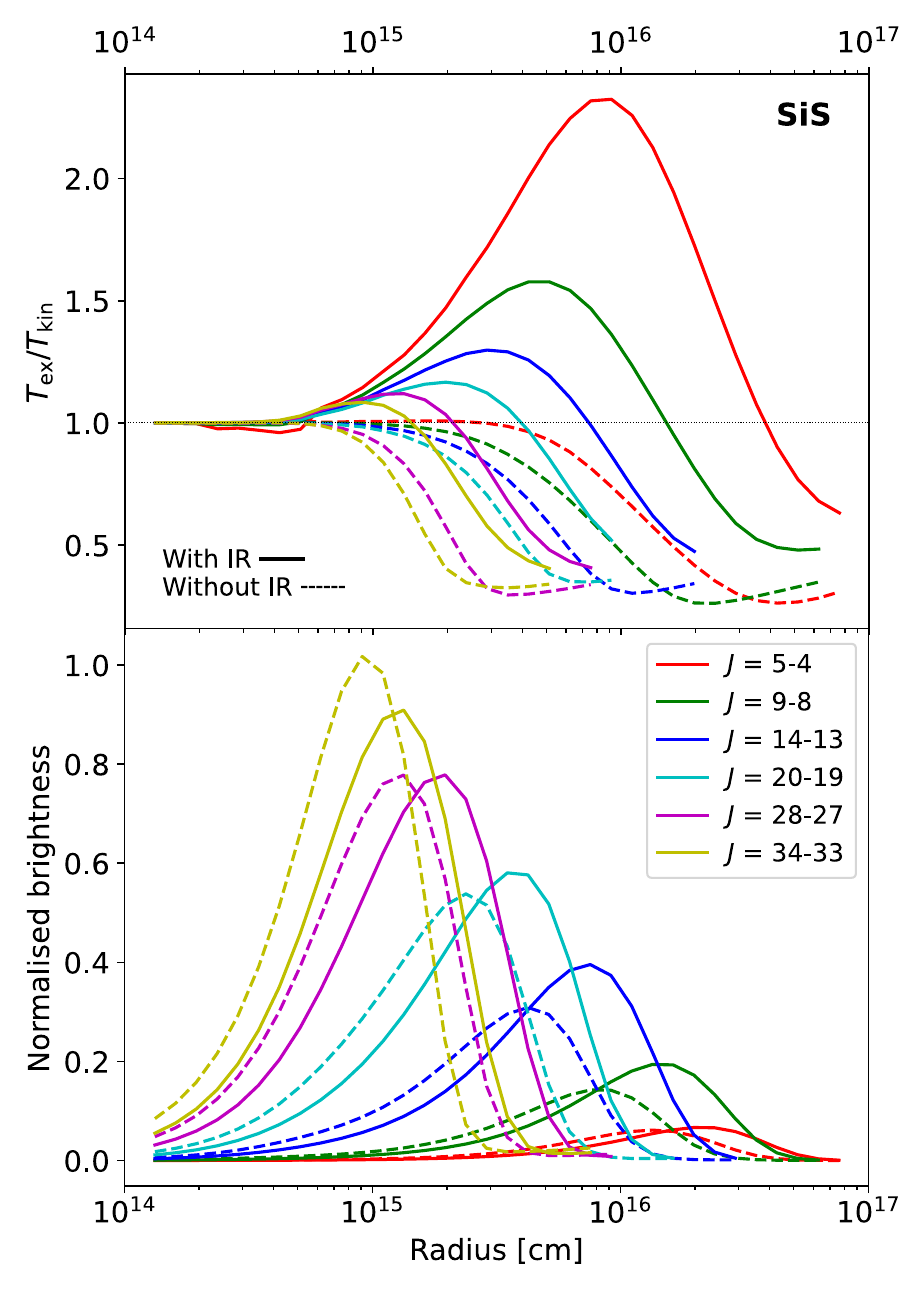}
    \caption{Same as Fig.~\ref{fig:sio_excitation_checks}, but for SiS.}
    \label{fig:sis_excitation_checks}
\end{figure}

\section{Discussion}
\label{sec:Discussion}
We derived the circumstellar fractional abundance profiles of SiO, SiS, and their {most abundant} isotopologues, for our sample of five carbon stars, by modelling the line emission of these species. Our RT models were constrained using both the radial distribution of the line emission, and information about the excitation conditions traced by different transitions, simultaneously, for all sources except AFGL 3068, for which we did not have interferometric emission maps available. We find models that fit the observations reasonably well for all sources, with a few exceptions as reported in Sect.~\ref{sec:Results}. We derived robust constraints on the SiO and SiS abundance profiles (see Table~\ref{tab:RT_modelling_results}, and Figs.~\ref{fig:SiO_chi_sq_maps}, \ref{fig:SiS_chi_sq_maps}) for all sources except AFGL 3068 (see Fig.~\ref{fig:AFGL_3068_chi_sq_maps}). Overall, our abundance estimates are consistent with previous works based on SD spectra \citep[e.g.][]{Agundez_et_al_2012, Fonfria_et_al_2015, Massalkhi_et_al_2019, Massalkhi_et_al_2024, Schoier_et_al_2007} and interferometric observations \citep[e.g.][]{Velilla-Prieto_et_al_2019, Schoier_et_al_2006}. Our poorly constrained SiS abundance profile for AFGL 3068 seems to underestimate the peak abundance and over-predict the $e$-folding radius, in comparison to SD SiS line modelling results by \citet{Schoier_et_al_2007}. For IRC~+10\,216, we find an SiO $R_\mathrm{e}$ $\sim$2.5 times larger than those reported by \citet{Massalkhi_et_al_2024} and \citet{Schoier_et_al_2006}. 

Our results show that the SiO abundances for IRAS 15194$-$5115, IRAS 15082$-$4808, and IRAS 07454$-$7112 are consistently larger than that of IRC~+10\,216, by a factor of $\sim$3--5, which is significant beyond the 1$\sigma$ uncertainties of the corresponding models (Table~\ref{tab:RT_modelling_results}). This is in line with the LTE modelling results from \citetalias{Unnikrishnan_et_al_2024}, indicating that simplistic LTE calculations are valid tools for deriving first-order estimates of circumstellar abundances and can reflect underlying trends.

We find that SiS has $\sim$2-15 times larger initial abundances than SiO (Table~\ref{tab:RT_modelling_results} and Fig.~\ref{fig:SiO_SiS_f0_vs_gas_density_comparison}), consistent with the findings of \citet{Massalkhi_et_al_2019, Massalkhi_et_al_2024} who performed RT modelling constrained by SD spectra. This is caused by the fact that the abundance of free O, required to form SiO, is typically low in carbon stars. Further, SiO in general has slightly larger $e$-folding radii than SiS (Table~\ref{tab:RT_modelling_results}), with both SiO and SiS typically having $R_\mathrm{e}$ smaller than that of CS \citepalias{Unnikrishnan_et_al_2025}. The SiO $R_\mathrm{e}$ values are $\sim$55-80\% of that of the CS ones, and the SiS $R_\mathrm{e}$ are $\sim$30-50\% of those of CS \citepalias[see][]{Unnikrishnan_et_al_2025}. The extents of the molecular emitting regions derived from the azimuthally averaged radial brightness profiles of CS, SiO, and SiS line emission from \citetalias{Unnikrishnan_et_al_2024} also follow the same order, i.e. $R_\mathrm{out, SiS}$ < $R_\mathrm{out, SiO}$ < $R_\mathrm{out, CS}$, where $R_\mathrm{out}$ denotes the outer radius of the emitting regions, though the observed radial extents of SiO and SiS emission are quite close. The circumstellar gas densities of our sample lie in the range 6.4$\times$10$^{-7}$ -- 3.0$\times$10$^{-6}$ $M_\odot$ yr$^{-1}$ km$^{-1}$ s. We note that for lower envelope densities (<$10^{-7}$ $M_\odot$ yr$^{-1}$ km$^{-1}$ s), \citet{Massalkhi_et_al_2024} found the radial extents of the abundance profiles of SiO to be larger than those of CS.

{\renewcommand{\arraystretch}{1.5}%
\begin{table*}[t]
   \caption{Isotopic ratios.}
   \label{tab:isotopic_ratios}
   \centering
   \begin{adjustbox}{width=14cm}
      \begin{tabular}{l l r r r r r r}
      \hline\hline & \\[-4ex]
      \makecell{\\Ratio} & \makecell{\\Species} & \multicolumn{5}{c}{Source}\\
      \cline{3-8} & \\[-3ex]
      & &  15194$-$5115 & 15082$-$4808 & 07454$-$7112 & AFGL 3068 & IRC~+10\,216 & Solar$^{(a)}$ \\
      \hline & \\[-3ex]
    $^{28}$Si/$^{29}$Si	&	SiO		&	$22.9^{+8.3}_{-12.4}$	 &	$21.4^{+8.7}_{-10.9}$	 &	$18.7^{+10.3}_{-8.4}$	 &  --  &	$22.0^{+15.4}_{-19.1}$	 &	20 \\
                        &	SiS		&	$24.2^{+10.7}_{-15.6}$	 &	$26.1^{+10.3}_{-17.7}$   &	$19.2^{+10.9}_{-9.8}$	 &  $16.0^{+\uparrow}_{-\downarrow}$  &	$21.0^{+11.6}_{-12.8}$	 &	   \\[0.5ex]
    
    $^{28}$Si/$^{30}$Si	&	SiO		&	$34.8^{+12.4}_{-19.2}$   &	$28.1^{+13.3}_{-11.0}$	 &  $27.5^{+15.4}_{-11.8}$	 &  --  &	$29.1^{+21.6}_{-20.1}$	 &	30   \\
                        &	SiS		&	--       &	$31.1^{+16.2}_{-16.9}$	 &  $30.0^{+16.6}_{-16.6}$	 &  $24.0^{+\uparrow}_{-\downarrow}$  &	$29.2^{+17.1}_{-16.2}$	 &	   \\[0.5ex]
    						    
    $^{29}$Si/$^{30}$Si	&	SiO		&	$1.5^{+0.7}_{-0.7}$ &	$1.3^{+0.7}_{-0.5}$ &	$1.4^{+0.7}_{-0.7}$ &  --  &	$1.3^{+1.1}_{-0.9}$ &	1.5 \\
    					&	SiS		&	--       &	$1.2^{+0.8}_{-0.5}$ &	$1.6^{+0.8}_{-0.9}$ &  $1.5^{+\uparrow}_{-\downarrow}$  &	$1.4^{+0.9}_{-0.7}$ &	   \\[0.5ex]

    $^{32}$S/$^{34}$S	&	SiS		&	$22.2^{+11.9}_{-12.0}$	 &	$24.6^{+11.9}_{-14.1}$	 &	$23.7^{+12.4}_{-15.2}$ 	 &  $18.0^{+\uparrow}_{-\downarrow}$  &	$20.7^{+12.4}_{-11.2}$   &  22 \\[1ex] \hline
      \end{tabular}
      \end{adjustbox}
      \tablefoot{Isotopic ratios are calculated as the ratio of the $f_0$ values of the best-fit models for the relevant species (Table~\ref{tab:RT_modelling_results}). The sub/super-scripts show the uncertainties on the values, and $^{+\uparrow}_{-\downarrow}$ indicates that the uncertainties have not been constrained. $^{(a)}$ \citet{Asplund_et_al_2009}.}
\end{table*}}

\subsection{Modified SiO abundance profiles}
\label{subsec:10216_SiO_fit_mismatch}
For the case of SiO around IRC~+10\,216, where we were not able to simultaneously fit the ALMA SiO $J = 2 - 1$ and the higher-$J$ SiO SD line profiles, we tried varying the inner abundance using a step function, as described in Sect.~\ref{subsec:sio_results}. This technique has been used in the literature \citep[e.g.][]{Schoier_et_al_2006, Schoier_et_al_2007, Danilovich_et_al_2019}, and works reasonably well in cases where the RT models are constrained either by the radial brightness distribution of a single spatially resolved line, or a small set of SD line profiles relatively close in excitation. However, as described in Sect.~\ref{subsec:sio_results}, as we have both spatial information and a large number of SD line profiles tracing different excitation conditions simultaneously available, there are additional constraints in place, particularly for the inner envelope, which is traced by both the small-aperture spectra from the ALMA data \citepalias[see Sect.~\ref{subsec:spectral_line_surveys} and][]{Unnikrishnan_et_al_2025} and the higher-$J$ SD lines. In our case, this leads to not being able to find good models using simple step functions that modify the abundance in the inner regions to fit either the SD line profiles or the small-aperture ALMA spectra, as any such modification to one affects the other as well. 

We note that it is indeed possible that the abundance profiles of these species can deviate from a smooth Gaussian, for example, due to depletion onto the dust given their refractory nature, or the presence of radial density variations \citep[see e.g.][\citetalias{Unnikrishnan_et_al_2024}]{Cordiner_and_Millar_2009}. The presence of a possible binary companion can also alter inner envelope abundances \citep[e.g.][]{Van_de_Sande_and_Millar_2022}. However, simplistic assumptions such as modifications to the inner abundance alone do not manage to model these effects when stringent observational constraints are placed on the RT models. Addressing these demands RT modelling, possibly in 3D, that properly takes into account not only the observed complex circumstellar density structures such as arcs and possible spirals \citepalias[see][]{Unnikrishnan_et_al_2024} and related variations in the gas and dust temperature profiles, but also includes detailed descriptions of the spectral emission from the central star and the dust radiation field, which is beyond the scope of this paper.

\subsection{A note on SiS J = 6 - 5}
\label{subsec:SiS_6_5_line}
As mentioned in Sect.~\ref{subsec:sis_results}, we can not find a model that fits the SiS $J = 5 - 4$ and $6 - 5$ lines simultaneously. This issue can be found in the literature as well. The modelled profiles of these two lines for IRC+10$\,$216 by \citet{Agundez_et_al_2012} underestimate the $J = 5 - 4$ line and marginally overestimate the $J = 6 - 5$ line, though in this case these modelled profiles are possibly within the uncertainty of the observed SD line profile. \citet{Velilla-Prieto_et_al_2019} used two different abundance profiles to fit the SiS $J = 5 - 4$ and $6 - 5$ lines for IRC~+10\,216. 

It is known from observations of IRC~+10\,216 that the SiS $J = 5 - 4$ and $J = 6 - 5$ lines display anti-correlated intensity variations with the stellar phase \citep[see][]{Carlstrom_et_al_1990}. It has also been suggested that there are possible IR overlaps between SiS $v = 1 - 0$ rovibrational transitions, some also involving the $J = 5$ level, and those of the $\nu_5$ mode of C$_2$H$_2$ and the $\nu_2$ mode of HCN \citep[see e.g.][]{Sahai_et_al_1984, Carlstrom_et_al_1990, Bieging_and_Tafalla_1993, Fonfria_Exposito_et_al_2006, Velilla-Prieto_et_al_2019}, which can alter the excitation of the $J = 5 - 4$ and $6 - 5$ lines. As a quick test, we set the Einstein A-coefficient ($A_{ij}$) of the SiS $v = 1 - 0,~J = 6 - 5$ ro-vibrational transition to zero, to check if this IR pumping route significantly affects the line intensities. We found that this causes significant changes, beyond typical calibration uncertainties, to both the modelled SiS $v = 0, J = 5 - 4$ and $6 - 5$ lines. The intensity of the SiS $J = 5 - 4$ line increased by $\sim$30\%, and that of the $J = 6 - 5$ decreased by $\sim$25\%. This is not to say that it is this specific ro-vibrational transition that is blended with those of other species as discussed above, but only to show that the $v = 0, J = 5$ level is highly sensitive to radiative excitation through the $v = 1$ levels, indicating that any possible blends that can affect the IR pumping schemes involving this level can significantly alter the line intensities of these $v = 0$ transitions. Such potential issues of overlaps with the transitions of other molecules in the infrared (IR) ro-vibrational levels (see Sect.~\ref{subsec:SiS_6_5_line}) cannot be directly taken into account in our RT models, which can only include transition information from the species being modelled. Since we find it possible to consistently model our $J = 5 - 4$ lines, and not the $J = 6 - 5$ lines, with the remaining SD lines across all sources, we chose to keep the $J = 5 - 4$ and discard the $J = 6 - 5$ from our modelling, in light of the above considerations.

\subsection{Radiative excitation}
\label{subsec:ir_pumping}
We tested whether radiative excitation via IR pumping \citep[see e.g.][\citetalias{Unnikrishnan_et_al_2025}]{Agundez_and_Cernicharo_2006, Agundez_et_al_2012, Velilla-Prieto_et_al_2019} contributes to the line excitation of SiO and SiS. For both molecules, we find that IR pumping significantly affects the line excitation and also the radial distribution of the emission, as also seen for CS in \citetalias{Unnikrishnan_et_al_2025}. The modelled line intensities drop by $\sim$30-60\% for SiO, and $\sim$30-50\% for SiS, when IR pumping is turned off. We note that dust emission is the main source of radiative excitation in these sources, and that the direct contribution of the stellar radiation is comparatively smaller, as found from varying the input stellar luminosities and dust opacities in Sect.~\ref{subsec:sio_results}. 

\begin{figure}[t]
    \centering
    \begin{subfigure}{0.425\textwidth}
        \includegraphics[width=\textwidth]{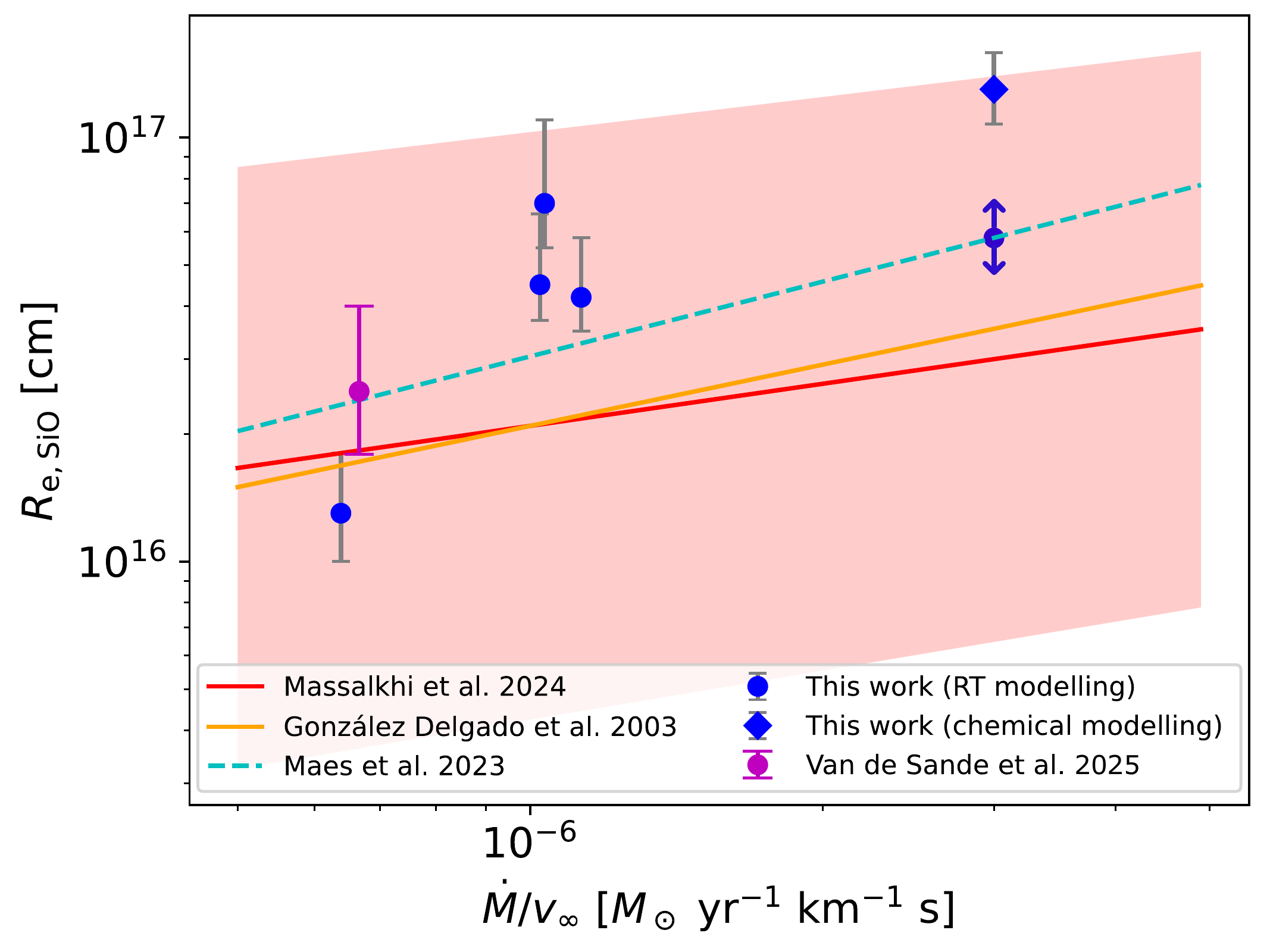}
        \caption{SiO}
        \label{subfig:Re_vs_gas_density_comparison_SiO}
    \end{subfigure}
    \par\bigskip
    \begin{subfigure}{0.425\textwidth}
        \includegraphics[width=\textwidth]{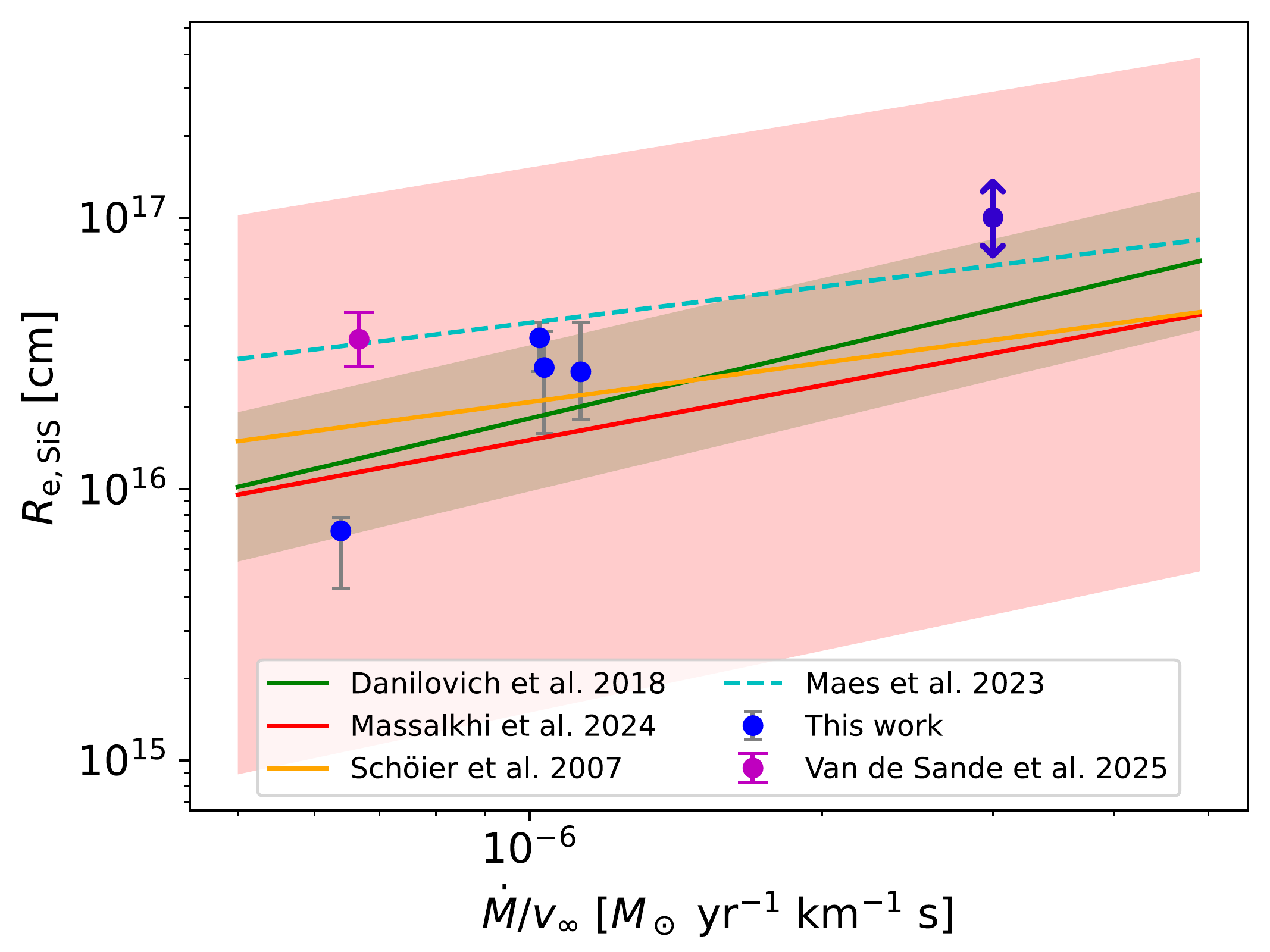}
        \caption{SiS}
        \label{subfig:Re_vs_gas_density_comparison_SiS}
    \end{subfigure}

    \caption{$R_\mathrm{e}$ of (a) SiO and (b) SiS abundance profiles versus the {mass-loss density (ratio of the gas MLR and terminal expansion velocity)}. The dashed cyan lines are from chemical modelling results \citep{Maes_et_al_2023}, while the other trends plotted are calculated from RT models. {The red and green shaded bands represent the uncertainties on the respective lines}. The magenta points are obtained from chemical modelling of a star of MLR 1.0$\times$10$^{−5}$ M$_\odot$ yr$^{−1}$, and a $\varv_\infty$ of 15 km s$^{−1}$, and show the error caused by the uncertainties on the kinetic data \citep{Van_de_Sande_et_al_submitted}. The rightmost blue circles in both plots are for AFGL 3068, and the open error bars indicate that the $R_\mathrm{e}$ is not well-constrained for this source. For SiO, the blue diamond indicates the $R_\mathrm{e}$ predicted by our chemical model (see Sect.~\ref{subsec:chemical_modelling}).}
    \label{fig:Re_vs_gas_density_comparisons}
\end{figure}

\begin{figure}[t]
    \centering
        \includegraphics[width=0.75\linewidth]{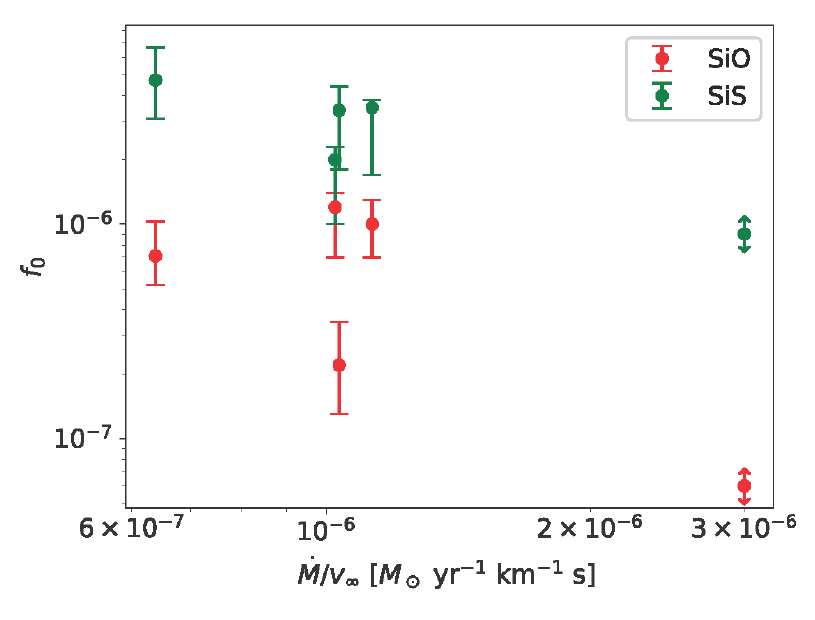}
    \caption{SiO and SiS peak abundances versus gas densities. The open error bars on the $f_0$ values for AFGL 3068, marked by arrows, indicate that the abundances for this star have not been well constrained.}
    \label{fig:SiO_SiS_f0_vs_gas_density_comparison}
\end{figure}

Figs.~\ref{fig:sio_excitation_checks} and \ref{fig:sis_excitation_checks} show the ratio of excitation temperature ($T_\mathrm{ex}$) to the gas kinetic temperature ($T_\mathrm{kin}$), and the line-emitting regions \citepalias[see][]{Unnikrishnan_et_al_2025} for different transitions for SiO and SiS, respectively, from the RT models with and without IR pumping taken into account. The peaks of the emitting regions are shifted inwards when IR pumping is turned off. Beyond the dense inner envelope where the level populations are thermalised ($T_\mathrm{ex}$ $\approx$ $T_\mathrm{kin}$), the line excitation becomes increasingly suprathermal ($T_\mathrm{ex}$ > $T_\mathrm{kin}$) as we move farther away from the star, as dust provides the IR radiation that radiatively excites the molecules. At the outer parts of the CSE, $T_\mathrm{ex}$ becomes lower than $T_\mathrm{kin}$ due to the decrease in gas density and the radiation field intensity.

\subsection{Isotopic ratios}
\label{subsec:isotopic_ratios}
We estimated the $^{28}$Si/$^{29}$Si, $^{28}$Si/$^{30}$Si, $^{29}$Si/$^{30}$Si, and $^{32}$S/$^{34}$S isotopic ratios from the derived abundances of SiO, SiS and their isotopologues (see Table~\ref{tab:isotopic_ratios}). The calculated ratios are very close to the solar values for all five sources, as in the case of the $^{32}$S/$^{34}$S ratios derived from CS abundances in \citetalias{Unnikrishnan_et_al_2025}. This is expected, as these elements and their isotopes are not produced by nucleosynthesis during the evolution of a star from the main-sequence to the AGB \citep[see][]{Karakas_and_Lugaro_2016}, except for the possible minor enhancement of $^{29}$Si and $^{30}$Si by the $s$-process during the AGB \citep[see e.g.][]{Zinner_et_al_2006, Decin_et_al_2010}. The proximity of these isotopic ratios to their solar values is also consistent with previous SD estimates \citep[see][]{Peng_et_al_2013, He_et_al_2008, Cernicharo_et_al_2000}. For IRAS 07454$-$7112, we find a Si$^{32}$S/Si$^{34}$S abundance ratio of 23.7 (Table~\ref{tab:isotopic_ratios}), which matches well with the solar $^{32}$S/$^{34}$S isotopic ratio of 22 \citep{Asplund_et_al_2009}, in contrast to the considerably lower Si$^{32}$S/Si$^{34}$S abundance ratio of 11.4 reported for this star by \citet{Danilovich_et_al_2018}, based on RT modelling of a limited number of SD spectra, including only the $J = 10 - 9$ and $11 - 10$ lines of Si$^{34}$S. 

We also note that the observed line intensity ratios involving the main isotopologues of SiO and SiS are consistently lower than their corresponding abundance ratios derived from the RT modelling. This is a consequence of the high optical depths in the principal isotopologue lines, as discussed in the case of CS in \citetalias{Unnikrishnan_et_al_2025}.

\begin{figure*}[t]
    \centering
    \includegraphics[width=0.9\linewidth]{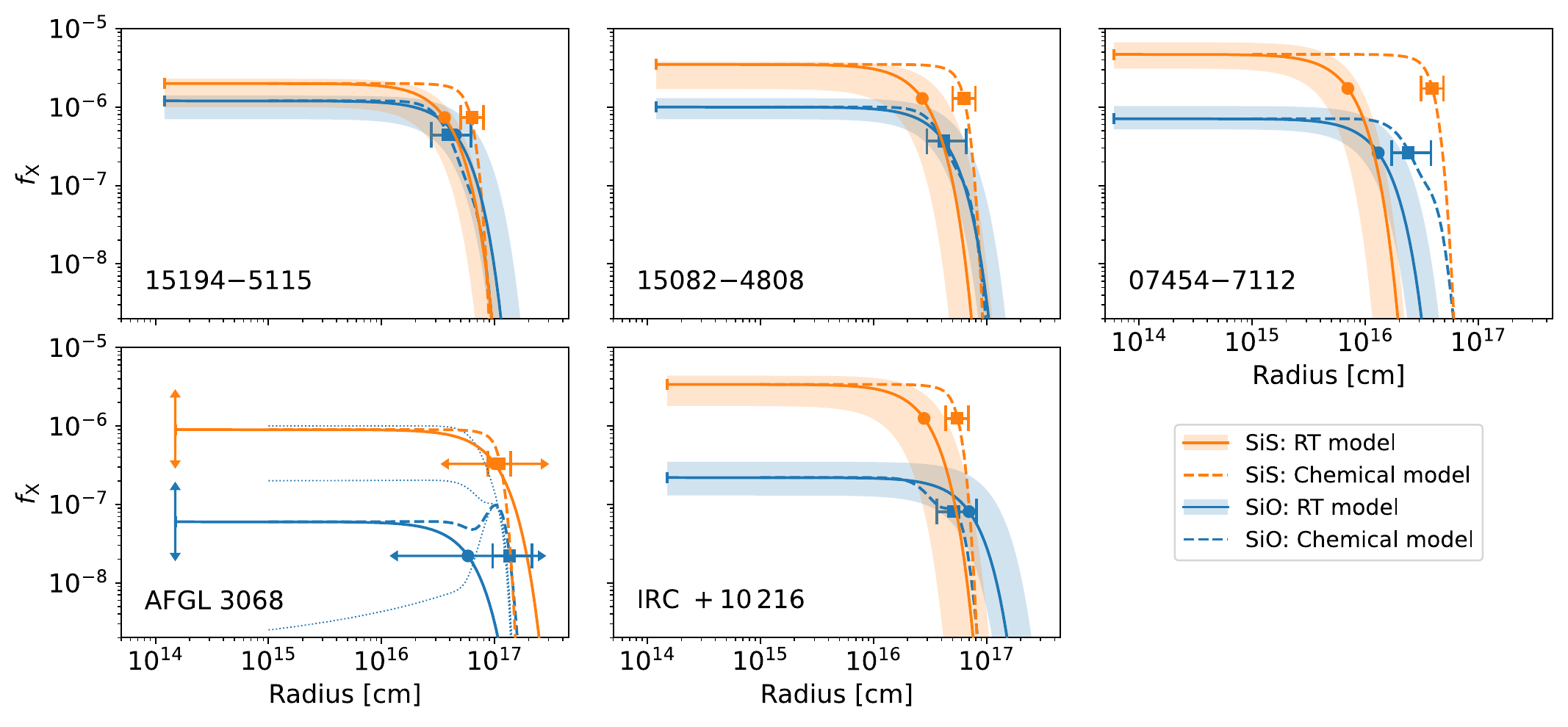}
    \caption{Comparison of abundance profiles obtained from RT modelling (solid lines with shaded 1$\sigma$ uncertainties) and chemical modelling (dashed lines) for SiO and SiS. The circles denote the $R_\mathrm{e}$ of the respective RT models and the squares mark the $R_\mathrm{e}$ of the chemical models. The arrows denote open error bars, indicating that the abundance profile has not been well constrained. The error bars on the chemical model $R_\mathrm{e}$ denote the estimated uncertainty in the chemical models based on reaction rate uncertainties \citep{Van_de_Sande_et_al_submitted}. The blue dotted lines for AFGL 3068 show chemical model abundance profiles with different initial SiO abundances {(1.0e-6, 2.0e-7, 2.5e-9)}.}
    \label{fig:CM_RT_abundance_comparisons}
\end{figure*}

\subsection{Trends with mass-loss density}
\label{subsec:trends_with_gas_density}
{The mass-loss density (ratio of gas MLR and the terminal gas expansion velocity) serves as a proxy for the circumstellar gas density in the outer regions of the CSE}. Our RT modelling results show that the $e$-folding radii of both SiO and SiS increase {with this parameter} (Fig.~\ref{fig:Re_vs_gas_density_comparisons}). This is in line with the empirical trends reported in the literature based on RT models constrained using SD lines \citep[e.g.][]{Gonzalez_Delgado_et_al_2003, Schoier_et_al_2006, Danilovich_et_al_2018, Massalkhi_et_al_2024} and chemical models \citep{Maes_et_al_2023}. Our SiO $R_\mathrm{e}$ values appear to follow a slightly sharper rise with density than the trends from the literature, while the increase in our SiS $R_\mathrm{e}$ with {this} density matches very well with previous estimates. {We note, however, that these observed trends in this work are influenced by the values for AFGL 3068, which are highly uncertain}.

The variation in peak abundances of the two species with {mass-loss} density is shown in Fig.~\ref{fig:SiO_SiS_f0_vs_gas_density_comparison}. The SiS peak abundances are consistently larger than those of SiO, as noted earlier. We find no clear correlation between the peak abundances of the two species. \citet{Schoier_et_al_2006} and \citet{Massalkhi_et_al_2019} found that SiO, and potentially also SiS abundances, decrease with increasing gas density in carbon star CSEs, based on SD observations of low-$J$ lines. \citet{Gonzalez_Delgado_et_al_2003} found a similar trend for SiO in M-type AGB stars, and \citet{Ramstedt_et_al_2009} tentatively found the same trend for S-type stars as well. This is possibly due to the increased adsorption of SiO and SiS onto dust grains at higher densities, caused by elevated collision rates. This indicates that these species, particularly SiO, may play an important role as a gas-phase precursor of dust in AGB stars. Unfortunately, our results cannot confirm or deny these reported trends.

\subsection{Chemical modelling}
\label{subsec:chemical_modelling}
We modelled the chemistry in the CSEs of our sources using a circumstellar chemical model\footnote{\url{https://github.com/MarieVdS/rate22_cse_code}} based on the \texttt{RATE22} update of the UMIST Database for Astrochemistry \citep[UDfA\footnote{\url{https://umistdatabase.uk}};][]{Millar_et_al_2024}. The chemical model used is the same as that in \citetalias{Unnikrishnan_et_al_2025}, a detailed description of which can be found in \citet{Millar_et_al_2000} and \citet{Van_de_Sande_et_al_2018_clumping_and_porosity}. We ran individual models for each of our sources, with the input physical parameters customised to match their circumstellar properties as modelled in \citetalias{Unnikrishnan_et_al_2025}, and set the initial abundances of SiO and SiS to those derived from our RT models for each source. The initial abundance of CS was obtained from \citetalias{Unnikrishnan_et_al_2025}, and those of other parent molecules were taken from \citet{Agundez_et_al_2020}. The resulting chemical model abundance profiles of SiO and SiS are shown in Fig.~\ref{fig:CM_RT_abundance_comparisons}, along with the corresponding abundance profiles derived from RT modelling. Including dust-gas chemistry in these models, to allow for possible changes in the radial gas-phase abundance distributions of our species due to dust-gas interactions, including depletion onto the dust, did not lead to significantly different abundance profiles. No significant depletion of the gas-phase species onto the dust occurs due to the relatively warm dust temperatures ($T_\mathrm{c}$ $\sim$ 1500-1800 K) that we employ as inputs to the RT models, based on our dust modelling results \citepalias{Unnikrishnan_et_al_2025}.

For SiO, it is seen that the abundance profiles predicted by the chemical model match very well with those from our RT models within the overall uncertainties of the models. This good match between the two abundance profiles shows that the chemical models do a reasonable job in reproducing the radial extent of SiO abundances in carbon star CSEs. This also indicates that our abundance profile for AFGL 3068, which is not well constrained (see Sect.~\ref{sec:Results}), is likely underestimating the SiO $e$-folding radius. To test this, we ran an SiO RT model for AFGL 3068 with the abundance profile predicted by the chemical model, which has a larger $R_\mathrm{e}$ than our original RT model, as can be seen from Fig.~\ref{fig:CM_RT_abundance_comparisons}, instead of a typical Gaussian profile. This also yielded reasonable fits to the observed SiO line profiles.

We note that the bump seen at $\sim$10$^{17}$ cm in the SiO chemical model abundance profile for AFGL 3068 is due to the reformation of SiO at large radii caused by reactions between H$_2$O and Si$^+$, which is produced from SiO by consecutive photoreactions. This bump becomes pronounced only when the initial abundance of SiO is sufficiently low {($\lesssim$ 2.0e-7)}, as can be seen from Fig.~\ref{fig:CM_RT_abundance_comparisons}. The same effect can also be seen to a smaller extent in the SiO profile for IRC~+10\,216, which displays a decrease in the profile at around 3$\times$10$^{16}$ cm, due to the photodissociation of SiO, followed by a small plateau, instead of a continuous decline, due to the above-mentioned reformation.

The chemical model over-predicts the SiS $e$-folding radii for all our sources, except AFGL 3068, where the observations provide a poor constraint (Sect.~\ref{subsubsec:sis_results_3068}). This indicates that, in general, SiS is photodissociated more inwards in the CSE than the chemical models predict. While the photodissociation rate of SiO is well known \citep{Heays_et_al_2017}, that of SiS remains a matter of uncertainty. The results from this work and other studies \citep{Danilovich_et_al_2018, Maes_et_al_2023, Van_de_Sande_et_al_submitted} imply a larger photodissociation rate for SiS, possibly similar to those of CS and SiO, than what is currently assumed in the chemical models. An accurate estimate of its destruction by interstellar photons is needed to fully test model predictions against observational data.

\section{Summary and conclusions}
\label{sec:Summary_and_conclusions}
We modelled the circumstellar SiO and SiS line emission of a sample of five C-type AGB stars, IRAS 15194$-$5115, IRAS 15082$-$4808, IRAS 07454$-$7112, AFGL 3068, and IRC~+10\,216, using detailed non-LTE radiative transfer modelling. The models were constrained by a combination of the radial brightness distributions from spatially resolved ALMA data, and a number of single-dish spectra from telescopes including APEX, \textit{Herschel}/HIFI, IRAM 30m, and the Yebes 40m, tracing a broad range of excitation conditions across the CSE. Though the heterogeneous origin of the datasets used to constrain the models introduces some uncertainty, the modelling overall yielded well-constrained, robust circumstellar fractional abundance profiles for both species and their {major} isotopologues, except for AFGL 3068, where we could not strictly constrain the abundance profiles due to a lack of spatially resolved observations. We confirmed that SiS has a larger peak abundance in carbon stars compared to SiO. For IRC+10216, our models cannot fit the SiO lower-$J$ ALMA spectra and the higher-$J$ SD spectra simultaneously. Introducing modified abundance profiles with higher abundance in the inner CSE did not solve this problem. We also investigated several other ways to address this issue without success, pointing to the limitations of 1D RT models based on simplistic assumptions about the stars and their circumstellar physical environments. 

We find that infrared pumping contributes significantly to the line excitation for both SiO and SiS. The derived SiS peak abundances are very similar across the sample, whereas the SiO abundances of the rest of the sources are a factor of $\sim$5 larger than that of IRC~+10\,216. The radial extent estimates of the SiO and SiS abundance profiles increase with circumstellar gas density as expected, {though the observed trends are limited by the large uncertainties on the derived values for AFGL 3068}. We compared our derived $e$-folding radii with those predicted by chemical models and found that while they match very well for SiO, the chemical models consistently overestimate the radial extent of SiS. We estimated the $^{28}$Si/$^{29}$Si, $^{28}$Si/$^{30}$Si, $^{29}$Si/$^{30}$Si, and $^{32}$S/$^{34}$S isotopic ratios for our sources, which correspond very well to the respective solar isotopic ratios, as expected.

These results point to the need for spatially resolved information, tracing a broad range of excitation conditions, to robustly constrain circumstellar molecular abundances. Radiative transfer modelling of many different molecular species, constrained by such observations, across a larger sample of AGB stars spanning a broad range of mass-loss rates, in particular sources with lower MLRs than the stars in our sample, complemented by updated chemical models, is needed to produce a comprehensive overview of circumstellar molecular chemistry in carbon stars.

\begin{acknowledgements}
The authors sincerely thank the anonymous referee for their constructive feedback, which improved the quality and clarity of this paper. The authors wish to thank P. Bergman for his support in setting up the ALI RT code. We thank M. Ag{\'u}ndez for providing Yebes 40m and IRAM 30m SiO and SiS line profiles of IRC~+10\,216, and J. H. Black for discussions about collisional rates and molecular data files. 

RU acknowledges data reduction support from the Nordic ALMA Regional Centre (ARC) node based at Onsala Space Observatory (OSO), Sweden. The Nordic ARC node is funded through Swedish Research Council grant No 2017-00648. We are grateful to D. Tafoya for his continuous support in ALMA data-related tasks.

EDB acknowledges financial support from the Swedish National Space Agency. 

MVdS acknowledges support from the Oort Fellowship at Leiden Observatory. 

TJM’s research at QUB is supported by grant ST/T000198/1 from the STFC.

TD is supported in part by the Australian Research Council through a Discovery Early Career Researcher Award (DE230100183).

MA acknowledges support from the Olle Engkvist Foundation under project 229-0368. 

SBC was supported by the NASA Planetary Science Division Internal Scientist Funding Program through the Fundamental Laboratory Research work package (FLaRe).

The work of MGR is supported by NOIRLab, which is managed by the Association of Universities for Research in Astronomy (AURA) under a cooperative agreement with the National Science Foundation.

This paper makes use of the following ALMA data: ADS/JAO.ALMA\#2013.1.00070.S, ADS/JAO.ALMA\#2015.1.01271.S, ADS/JAO.ALMA\#2017.1.00595.S, ADS/JAO.ALMA\#2013.1.00432.S. ALMA is a partnership of ESO (representing its member states), NSF (USA) and NINS (Japan), together with NRC (Canada), MOST and ASIAA (Taiwan), and KASI (Republic of Korea), in cooperation with the Republic of Chile. The Joint ALMA Observatory is operated by ESO, AUI/NRAO and NAOJ. 

This paper is based on observations with the Atacama Pathfinder EXperiment (APEX) telescope. APEX is a collaboration between the Max Planck Institute for Radio Astronomy, the European Southern Observatory, and the Onsala Space Observatory. Swedish observations on APEX are supported through Swedish Research Council grant No 2017-00648. The APEX observations were obtained under project numbers O-0107.F-9310 (SEPIA/B5), O-0104.F-9305 (PI230), and O-087.F-9319, O-094.F-9318, O-096.F-9336, and O-098.F-9303 (SHeFI). 

HIFI has been designed and built by a consortium of institutes and university departments from across Europe, Canada, and the United States (NASA) under the leadership of SRON, Netherlands Institute for Space Research, Groningen, The Netherlands, and with major contributions from Germany, France and the US. Consortium members are Canada: CSA, U. Waterloo; France: CESR, LAB, LERMA, IRAM; Germany: KOSMA, MPIfR, MPS; Ireland: NUI Maynooth; Italy: ASI, IFSI-INAF, Osservatorio Astrofísico di Arcetri-INAF; The Netherlands: SRON, TUD; Poland: CAMK, CBK; Spain: Observatorio Astronómico Nacional (IGN), Centro de Astrobiología (INTA-CSIC); Sweden: Chalmers University of Technology – MC2, RSS \& GARD, Onsala Space Observatory, Swedish National Space Board, Stockholm University – Stockholm Observatory; Switzerland: ETH Zurich, FHNW; USA: CalTech, JPL, NHSC. 

This work is based on observations carried out with the IRAM 30m and the Yebes 40m telescopes. IRAM is supported by INSU/CNRS (France), MPG (Germany) and IGN (Spain). The Yebes 40m telescope at Yebes Observatory is operated by the Spanish Geographic Institute (IGN, Ministerio de Transportes, Movilidad y Agenda Urbana).

This research has made use of NASA’s Astrophysics Data System (ADS, \url{https://ui.adsabs.harvard.edu}). 
 
This work made use of Astropy (\url{https://www.astropy.org}), a community-developed core Python package and an ecosystem of tools and resources for astronomy \citep{astropy:2013, astropy:2018, astropy:2022}. 

This work has made use of GILDAS (\url{http://www.iram.fr/IRAMFR/GILDAS/}) and CASA (\url{https://casa.nrao.edu}) software to reduce and analyse data. 

The computations in this work were run in the Vera HPC cluster using resources provided by C3SE, the Chalmers e-Commons e-Infrastructure group at Chalmers University of Technology, Gothenburg, Sweden.
\end{acknowledgements}

\bibliographystyle{aa}
\bibliography{references}

\begin{appendix}
\onecolumn

\section{Observed line intensities}
\label{app:Appendix_A}

\begin{table*}[h]
   \caption{SiO lines used in this work.}
   \label{tab:SiO_line_intensities}
   \centering
      \begin{adjustbox}{width=18cm}
      \begin{tabular}{c r c c c c c c c}
      \hline\hline & \\[-2ex]
      \makecell{Transition} & \makecell{Rest Frequency\\\ [GHz]} & \makecell{E$_{up}$\\\ [K]} & \makecell{Telescope} & \multicolumn{5}{c}{\makecell{Integrated Intensity [Jy km s$^{-1}$ for ALMA and ACA; K km s$^{-1}$ for others]}} \\
      \cline{5-9} & \\[-2ex]  
      & & & & 15194$-$5115 & 15082$-$4808 & 07454$-$7112 & AFGL 3068 & IRC~+10\,216 \\
      \hline & \\[-2ex]
      $1 - 0$ & 43.423853 & 2.08 & Yebes & $-$ & $-$ & $-$ & $-$ & 15.4$^{(b)}$\\\\[-3mm]

      $2 - 1$ & 86.846985 & 6.25 & ALMA & 96.1 & 35.1 & 15.8 & $-$ & 422.5$^{(c)}$\\
             &             &      & IRAM & $-$ & $-$ & $-$ & $-$ & 65.2$^{(d)}$\\\\[-3mm]

      $3 - 2$ & 130.268683 & 12.50 & IRAM & $-$ & $-$ & $-$ & $-$ & 139.4$^{(d)}$\\\\[-3mm]

      $4 - 3$ & 173.688238 & 20.84 & APEX & 12.4 & 1.2 & 2.4 & 0.9 & 62.1\\
              &             &       & IRAM & $-$ & $-$ & $-$ & $-$ & 226.4$^{(d)}$\\\\[-3mm]

      $5 - 4$ & 217.104919 & 31.26 & APEX & 15.2 & 1.2 & 3.6 & 0.9 & 75.4\\
              &             &       & IRAM & $-$ & $-$ & $-$ & $-$ & 248.2$^{(d)}$\\
              &             &       & ACA & $-$ & $-$ & 97.5$^{(a)}$ & 23.1$^{(a)}$ & $-$\\\\[-3mm]

      $6 - 5$ & 260.518009 & 43.76 & APEX & 13.6 & 1.2 & 4.2 & 0.9 & 89.9\\
              &             &       & IRAM & $-$ & $-$ & $-$ & $-$ & 281.4$^{(d)}$\\\\[-3mm]

      $7 - 6$ & 303.926812 & 58.35 & APEX & 13.8 & $-$ & $-$ & $-$ & $-$\\
              &             &       & IRAM & $-$ & $-$ & $-$ & $-$ & 349.9$^{(d)}$\\\\[-3mm]

      $8 - 7$ & 347.330581 & 75.02 & APEX & 14.4 & $-$ & $-$ & $-$ & $-$\\
              &             &       & IRAM & $-$ & $-$ & $-$ & $-$ & 380.6$^{(d)}$\\\\[-3mm]

      $12 - 11$ & 520.881187 & 162.52 & HIFI & 1.4 & $-$ & $-$ & $-$ & $-$\\
      $13 - 12$ & 564.249098 & 189.60 & HIFI & 1.3 & $-$ & $-$ & $-$ & $-$\\
      $14 - 13$ & 607.607719 & 218.76 & HIFI & 1.3 & $-$ & $-$ & $-$ & $-$\\
      $15 - 14$ & 650.956290 & 250.00 & HIFI & 1.2 & $-$ & $-$ & $-$ & $-$\\
      $16 - 15$ & 694.294114 & 283.32 & HIFI & 0.9 & $-$ & $-$ & $-$ & $-$\\
      $18 - 17$ & 780.934648 & 356.20 & HIFI & 0.5 & $-$ & $-$ & $-$ & $-$\\
      $19 - 18$ & 824.235900 & 395.75 & HIFI & 0.4 & $-$ & $-$ & $-$ & $-$\\
      $20 - 19$ & 867.523546 & 437.39 & HIFI & 0.4 & $-$ & $-$ & $-$ & $-$\\
      $21 - 20$ & 910.796851 & 481.10 & HIFI & 0.8 & $-$ & $-$ & $-$ & $-$\\
      $22 - 21$ & 954.055103 & 526.89 & HIFI & 0.6 & $-$ & $-$ & $-$ & $-$\\
      \hline
      \end{tabular}
      \end{adjustbox}
      \tablefoot{The ALMA and APEX lines reported for IRAS 15194$-$5115, IRAS 15082$-$4808, and IRAS 07454$-$7112 are from \citetalias{Unnikrishnan_et_al_2024}, unless otherwise specified. All HIFI lines listed are from our \textit{Herschel}/HIFI spectral survey of IRAS 15194$-$5115 \citepalias[see][]{Unnikrishnan_et_al_2024}. For the ALMA (and ACA) lines, the integrated intensity values reported are in units of Jy km s$^{-1}$ and are calculated using spectra extracted from apertures large enough to encompass all detected line emission. For all other SD observations (APEX, HIFI, IRAM, Yebes), the integrated intensities are given in units of K km s$^{-1}$, in the main beam ($T_\mathrm{MB}$) temperature scale. The relevant beam size ranges and main beam efficiencies for the SD telescopes are described in \citetalias{Unnikrishnan_et_al_2025}, and the individual beam sizes at each transition frequency are indicated in Fig.~\ref{fig:15194-5115_SiO} and the relevant figures in Appendix~\ref{app:Appendix_B}, alongside the corresponding line spectra. $^{(a)}$ The DEATHSTAR ACA survey \citep{Ramstedt_et_al_2020, Andriantsaralaza_et_al_2021}; $^{(b)}$ \citet{Massalkhi_et_al_2024}; $^{(c)}$ \citet{Velilla-Prieto_et_al_2019}; $^{(d)}$ \citet{Agundez_et_al_2012}.}
\end{table*}

\begin{table*}[ht]
   \caption{$^{29}$SiO lines used in this work.}
   \label{tab:29SiO_line_intensities}
   \centering
      \begin{adjustbox}{width=18cm}
      \begin{tabular}{c r c c c c c c c}
      \hline\hline & \\[-2ex]
      \makecell{Transition} & \makecell{Rest Frequency\\\ [GHz]} & \makecell{E$_{up}$\\\ [K]} & \makecell{Telescope} & \multicolumn{5}{c}{\makecell{Integrated Intensity [Jy km s$^{-1}$ for ALMA and ACA; K km s$^{-1}$ for others]}} \\
      \cline{5-9} & \\[-2ex]  
      & & & & 15194$-$5115 & 15082$-$4808 & 07454$-$7112 & AFGL 3068 & IRC~+10\,216 \\
      \hline & \\[-2ex]
      $2 - 1$ & 85.759199 & 6.17 & ALMA & 9.8 & 2.5 & 1.3 & $-$ & 31.1$^{(c)}$\\\\[-3mm]
      $4 - 3$ & 171.512802 & 20.58 & APEX & 1.2 & 0.5 & 0.3 & $-$ & 5.4\\
      $5 - 4$ & 214.385757 & 30.87 & APEX & 1.6 & 0.5 & 0.4 & $-$ & 7.1\\
      $6 - 5$ & 257.255215 & 43.21 & APEX & 1.5 & 0.5 & 0.4 & $-$ & 6.9\\
      $7 - 6$ & 300.120477 & 57.62 & APEX & 1.3 & $-$ & $-$ & $-$ & $-$\\
      $8 - 7$ & 342.980842 & 74.08 & APEX & 1.1 & $-$ & $-$ & $-$ & $-$\\
              &            &       & ACA  & $-$ & $-$ & 17.5$^{(a)}$ & $-$ & $-$\\\\[-3mm]
      $12 - 11$ & 514.359357 & 160.48 & HIFI & 0.1 & $-$ & $-$ & $-$ & $-$\\
      \hline
      \end{tabular}
      \end{adjustbox}
      \tablefoot{As in Table.~\ref{tab:SiO_line_intensities}, but for $^{29}$SiO.}
\end{table*}

\begin{table*}[ht]
   \caption{$^{30}$SiO lines used in this work.}
   \label{tab:30SiO_line_intensities}
   \centering
      \begin{adjustbox}{width=17.25cm}
      \begin{tabular}{c r c c c c c c c}
      \hline\hline & \\[-2ex]
      \makecell{Transition} & \makecell{Rest Frequency\\\ [GHz]} & \makecell{E$_{up}$\\\ [K]} & \makecell{Telescope} & \multicolumn{5}{c}{\makecell{Integrated Intensity [Jy km s$^{-1}$ for ALMA and ACA; K km s$^{-1}$ for others]}} \\
      \cline{5-9} & \\[-2ex]  
      & & & & 15194$-$5115 & 15082$-$4808 & 07454$-$7112 & AFGL 3068 & IRC~+10\,216 \\
      \hline & \\[-2ex]
      $2 - 1$ & 169.486877 & 20.34 & ALMA & $-$ & $-$ & $-$ & $-$ & 23.1$^{(c)}$\\\\[-3mm]
      $4 - 3$ & 169.486877 & 20.34 & APEX & 0.7 & 0.3 & 0.2 & $-$ & 3.7\\
      $5 - 4$ & 211.853474 & 30.50 & APEX & 1.5 & 0.6 & 0.5 & $-$ & 13.2\\
      $6 - 5$ & 254.216656 & 42.70 & APEX & 1.0 & 0.4 & 0.3 & $-$ & 5.5\\
      $7 - 6$ & 296.575740 & 56.94 & APEX & 0.9 & $-$ & $-$ & $-$ & $-$\\
      $8 - 7$ & 338.930044 & 73.20 & APEX & 1.1 & $-$ & $-$ & $-$ & $-$\\
      \hline
      \end{tabular}
      \end{adjustbox}
      \tablefoot{As in Table.~\ref{tab:SiO_line_intensities}, but for $^{30}$SiO.}
\end{table*}

\begin{table*}[ht]
   \caption{SiS lines used in this work.}
   \label{tab:SiS_line_intensities}
   \centering
      \begin{adjustbox}{width=17.25cm}
      \begin{tabular}{c r c c c c c c c}
      \hline\hline & \\[-2ex]
      \makecell{Transition} & \makecell{Rest Frequency\\\ [GHz]} & \makecell{E$_{up}$\\\ [K]} & \makecell{Telescope} & \multicolumn{5}{c}{\makecell{Integrated Intensity [Jy km s$^{-1}$ for ALMA and ACA; K km s$^{-1}$ for others]}} \\
      \cline{5-9} & \\[-2ex]  
      & & & & 15194$-$5115 & 15082$-$4808 & 07454$-$7112 & AFGL 3068 & IRC~+10\,216 \\
      \hline & \\[-2ex]
      $2 - 1$   & 36.309627  & 2.61   & Yebes & $-$ & $-$ & $-$ & $-$ & 9.1$^{(b)}$\\\\[-3mm]

      $5 - 4$   & 90.771564  & 13.07  & ALMA & 43.8 & 23.1 & 12.1 & $-$ & 645.4$^{(c)}$, 673.0$^{(e)}$\\
                &            &        & IRAM & $-$ & $-$ & $-$ & $-$ & 121.3$^{(d)}$\\\\[-3mm]

      $6 - 5$   & 108.924301 & 18.30  & ALMA & 45.8 & 27.0 & 14.9 & $-$ & 653.7$^{(c)}$\\
                &            &        & IRAM & $-$ & $-$ & $-$ & $-$ & 116.2$^{(d)}$\\\\[-3mm]

      $8 - 7$   & 145.227053 & 31.37  & IRAM & $-$ & $-$ & $-$ & $-$ & 289.3$^{(d)}$\\\\[-3mm]

      $9 - 8$   & 163.376785 & 39.21  & APEX & 5.2 & 3.8 & 2.3 & 2.4 & 96.7\\
                &            &        & IRAM & $-$ & $-$ & $-$ & $-$ & 370.0$^{(d)}$\\\\[-3mm]
                
      $10 - 9$  & 181.525218 & 47.92  & APEX & 4.9 & 2.9 & 2.0 & 1.9 & 83.5\\
                &            &        & IRAM & $-$ & $-$ & $-$ & $-$ & 272.0$^{(d)}$\\\\[-3mm]
                
      $11 - 10$ & 199.672229 & 57.50  & APEX & 5.2 & 4.1 & 2.4 & 2.2 & 99.7\\
                &            &        & IRAM & $-$ & $-$ & $-$ & $-$ & 368.8$^{(d)}$\\\\[-3mm]
                
      $12 - 11$ & 217.817663 & 67.95  & APEX & 5.2 & 3.9 & 3.1 & 3.3 & 114.7\\
                &            &        & IRAM & $-$ & $-$ & $-$ & $-$ & 302.8$^{(d)}$\\
                &            &        & ACA  & $-$ & $-$ & 85.3$^{(a)}$ & 89.6$^{(a)}$ & $-$\\\\[-3mm]
                
      $13 - 12$ & 235.961363 & 79.28  & APEX & 8.7 & 4.6 & 3.9 & 3.5 & 136.4\\
                &            &        & IRAM & $-$ & $-$ & $-$ & $-$ & 406.6$^{(d)}$\\\\[-3mm]
                
      $14 - 13$ & 254.103210 & 91.47  & APEX & 7.3 & 4.1 & 4.7 & 3.1 & 137.4\\
                &            &        & IRAM & $-$ & $-$ & $-$ & $-$ & 591.4$^{(d)}$\\\\[-3mm]
                
      $15 - 14$ & 272.243052 & 104.54 & APEX & 7.2 & $-$ & $-$ & $-$ & $-$\\
                &            &        & IRAM & $-$ & $-$ & $-$ & $-$ & 654.8$^{(d)}$\\\\[-3mm]
                
      $16 - 15$ & 290.380757 & 118.47 & APEX & 8.8 & $-$ & 5.0$^{(f)}$ & $-$ & $-$\\
                &            &        & IRAM & $-$ & $-$ & $-$ & $-$ & 613.2$^{(d)}$\\\\[-3mm]
                
      $17 - 16$ & 308.516143 & 133.28 & APEX & 5.5 & $-$ & $-$ & $-$ & $-$\\
                &            &        & IRAM & $-$ & $-$ & $-$ & $-$ & 381.8$^{(d)}$\\\\[-3mm]
                
      $18 - 17$ & 326.649109 & 148.96 & APEX & 3.7 & $-$ & $-$ & $-$ & $-$\\
                &            &        & IRAM & $-$ & $-$ & $-$ & $-$ & 522.7$^{(d)}$\\\\[-3mm]
                
      $19 - 18$ & 344.779481 & 165.50 & APEX & 5.7 & $-$ & 7.7$^{(f)}$ & $-$ & $-$\\
                &            &        & IRAM & $-$ & $-$ & $-$ & $-$ & 471.5$^{(d)}$\\\\[-3mm]

      $20 - 19$ & 362.907164 & 182.92 & APEX & 5.3 & $-$ & $-$ & $-$ & $-$\\\\[-3mm]
      
      $29 - 28$ & 525.909965 & 378.80 & HIFI & 0.4 & $-$ & $-$ & $-$ & $-$\\
      $30 - 29$ & 544.002518 & 404.90 & HIFI & 0.4 & $-$ & $-$ & $-$ & $-$\\
      $31 - 30$ & 562.090777 & 431.88 & HIFI & 0.3 & $-$ & $-$ & $-$ & $-$\\
      $32 - 31$ & 580.174604 & 459.72 & HIFI & 0.6 & $-$ & $-$ & $-$ & $-$\\
      $33 - 32$ & 598.253828 & 488.43 & HIFI & 0.4 & $-$ & $-$ & $-$ & $-$\\
      $34 - 33$ & 616.328337 & 518.01 & HIFI & 0.6 & $-$ & $-$ & $-$ & $-$\\
      $35 - 34$ & 634.397967 & 548.46 & HIFI & 0.7 & $-$ & $-$ & $-$ & $-$\\
      $36 - 35$ & 652.462584 & 579.77 & HIFI & 0.6 & $-$ & $-$ & $-$ & $-$\\
      $37 - 36$ & 670.522052 & 611.95 & HIFI & 0.4 & $-$ & $-$ & $-$ & $-$\\
      $38 - 37$ & 688.576198 & 645.00 & HIFI & 0.3 & $-$ & $-$ & $-$ & $-$\\
      $39 - 38$ & 706.624908 & 678.91 & HIFI & 0.4 & $-$ & $-$ & $-$ & $-$\\
      $42 - 41$ & 760.736942 & 785.84 & HIFI & 0.6 & $-$ & $-$ & $-$ & $-$\\
      \hline
      \end{tabular}
      \end{adjustbox}
      \tablefoot{As in Table.~\ref{tab:SiO_line_intensities}, but for SiS. $^{(e)}$ ALMA project 2015.1.01271.S (PI: D. Keller), $^{(f)}$ \citet{Danilovich_et_al_2018}.}
\end{table*}

\begin{table*}[ht]
   \caption{$^{29}$SiS lines used in this work.}
   \label{tab:29SiS_line_intensities}
   \centering
      \begin{adjustbox}{width=18cm}
      \begin{tabular}{c r c c c c c c c}
      \hline\hline & \\[-2ex]
      \makecell{Transition} & \makecell{Rest Frequency\\\ [GHz]} & \makecell{E$_{up}$\\\ [K]} & \makecell{Telescope} & \multicolumn{5}{c}{\makecell{Integrated Intensity [Jy km s$^{-1}$ for ALMA and ACA; K km s$^{-1}$ for others]}} \\
      \cline{5-9} & \\[-2ex]  
      & & & & 15194$-$5115 & 15082$-$4808 & 07454$-$7112 & AFGL 3068 & IRC~+10\,216 \\
      \hline & \\[-2ex]
      $5 - 4$ & 89.103749 & 12.83 & ALMA & 2.5 & 1.7 & 0.4 & $-$ & 23.9$^{(c)}$\\
      $6 - 5$ & 106.922980 & 17.96 & ALMA & 3.5 & $-$ & 1.1 & $-$ & 48.9$^{(c)}$\\\\[-3mm]
      
      $9 - 8$ & 160.375151 & 38.49 & APEX & $-$ & $-$ & $-$ & 0.2 & 6.9\\
      $10 - 9$ & 178.190241 & 47.04 & APEX & $-$ & $-$ & $-$ & $-$ & 6.6\\
      $11 - 10$ & 196.003949 & 56.44 & APEX & $-$ & 0.3 & $-$ & $-$ & $-$\\
      $12 - 11$ & 213.816140 & 66.71 & APEX & $-$ & 0.2 & 0.2 & 0.3 & 8.6\\
      $13 - 12$ & 231.626673 & 77.82 & APEX & $-$ & 0.2 & 0.3 & 0.3 & 10.1\\
                &            &       & ACA  & $-$ & $-$ & 8.2$^{(a)}$ & 8.3$^{(a)}$ & $-$\\\\[-3mm]
                
      $14 - 13$ & 249.435412 & 89.79 & APEX & $-$ & 0.2 & 0.3 & 0.3 & 9.6\\
      $15 - 14$ & 267.242218 & 102.62 & APEX & $-$ & $-$ & 0.4 & 0.4 & 10.6\\
      \hline
      \end{tabular}
      \end{adjustbox}
      \tablefoot{As in Table.~\ref{tab:SiO_line_intensities}, but for $^{29}$SiS.}
\end{table*}

\begin{table*}[ht]
   \caption{$^{30}$SiS lines used in this work.}
   \label{tab:30SiS_line_intensities}
   \centering
      \begin{adjustbox}{width=18cm}
      \begin{tabular}{c r c c c c c c c}
      \hline\hline & \\[-2ex]
      \makecell{Transition} & \makecell{Rest Frequency\\\ [GHz]} & \makecell{E$_{up}$\\\ [K]} & \makecell{Telescope} & \multicolumn{5}{c}{\makecell{Integrated Intensity [Jy km s$^{-1}$ for ALMA and ACA; K km s$^{-1}$ for others]}} \\
      \cline{5-9} & \\[-2ex]  
      & & & & 15194$-$5115 & 15082$-$4808 & 07454$-$7112 & AFGL 3068 & IRC~+10\,216 \\
      \hline & \\[-2ex]
      $5 - 4$ & 87.550558 & 12.61 & APEX & $-$ & $-$ & $-$ & $-$ & 20.6$^{(c)}$\\
      $6 - 5$ & 105.059203 & 17.65 & APEX & $-$ & $-$ & 0.6 & $-$ & 31.3$^{(c)}$\\
      $10 - 9$ & 175.084456 & 46.22 & APEX & $-$ & $-$ & $-$ & $-$ & 4.5\\
      $11 - 10$ & 192.587770 & 55.46 & APEX & $-$ & $-$ & $-$ & $-$ & 5.1\\
      $12 - 11$ & 210.089618 & 65.54 & APEX & $-$ & 0.1 & 0.2 & 0.2 & 6.3\\
      $13 - 12$ & 227.589867 & 76.47 & APEX & $-$ & 0.2 & 0.2 & 0.2 & 6.4\\
      $14 - 13$ & 245.088383 & 88.23 & APEX & $-$ & $-$ & 0.1 & 0.2 & 5.5\\
      $15 - 14$ & 262.585034 & 100.83 & APEX & $-$ & $-$ & 0.2 & 0.2 & 5.5\\\\[-3mm]
      
      $19 - 18$ & 332.550309 & 159.63 & ACA  & $-$ & $-$ & 12.4$^{(a)}$ & 9.7$^{(a)}$ & $-$\\
      \hline
      \end{tabular}
      \end{adjustbox}
      \tablefoot{As in Table.~\ref{tab:SiO_line_intensities}, but for $^{30}$SiS.}
\end{table*}

\begin{table*}[ht]
   \caption{Si$^{34}$S lines used in this work.}
   \label{tab:Si34S_line_intensities}
   \centering
      \begin{adjustbox}{width=18cm}
      \begin{tabular}{c r c c c c c c c}
      \hline\hline & \\[-2ex]
      \makecell{Transition} & \makecell{Rest Frequency\\\ [GHz]} & \makecell{E$_{up}$\\\ [K]} & \makecell{Telescope} & \multicolumn{5}{c}{\makecell{Integrated Intensity [Jy km s$^{-1}$ for ALMA and ACA; K km s$^{-1}$ for others]}} \\
      \cline{5-9} & \\[-2ex]  
      & & & & 15194$-$5115 & 15082$-$4808 & 07454$-$7112 & AFGL 3068 & IRC~+10\,216 \\
      \hline & \\[-2ex]
      $5 - 4$ & 88.285828 & 12.71 & ALMA & $-$ & 1.0 & 0.4 & $-$ & 19.0$^{(c)}$\\
      $6 - 5$ & 105.941503 & 17.80 & ALMA & 1.9 & $-$ & 0.7 & $-$ & 34.6$^{(c)}$\\\\[-3mm]
      
      $9 - 8$ & 158.903106 & 38.13 & APEX & $-$ & $-$ & $-$ & $-$ & 6.2\\
      $10 - 9$ & 176.554715 & 46.61 & APEX & $-$ & $-$ & $-$ & $-$ & 6.0\\
      $11 - 10$ & 194.204969 & 55.93 & APEX & $-$ & $-$ & $-$ & 0.2 & 7.5\\
      $13 - 12$ & 229.500868 & 77.11 & APEX & $-$ & 0.3 & 0.2 & 0.2 & 8.8\\
      $14 - 13$ & 247.146242 & 88.97 & APEX & $-$ & 0.2 & 0.3 & 0.3 & 8.4\\
      $15 - 14$ & 264.789719 & 101.68 & APEX & $-$ & 0.3 & 0.3 & 0.3 & 8.9\\
      $16 - 15$ & 282.431163 & 115.23 & APEX & 0.5 & $-$ & $-$ & $-$ & $-$\\
      \hline
      \end{tabular}
      \end{adjustbox}
      \tablefoot{As in Table.~\ref{tab:SiO_line_intensities}, but for Si$^{34}$S.}
\end{table*}

\clearpage

\section{SiO and SiS line fits}
\label{app:Appendix_B}
\FloatBarrier


\begin{figure*}[h]
    \centering
    \includegraphics[width=0.95\linewidth]{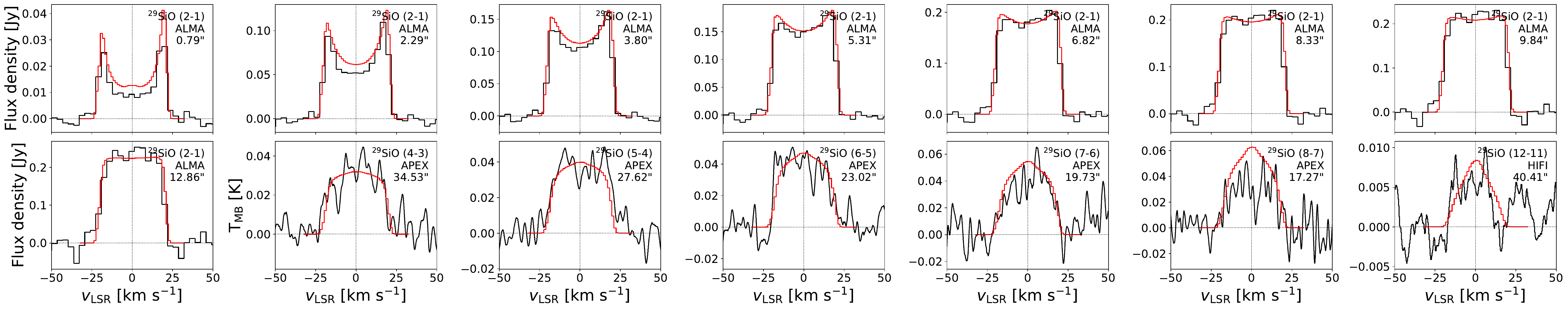}
    \caption{Observed (black) and modelled (red) $^{{29}}$SiO line profiles for IRAS 15194-5115. The transition quantum numbers, telescope used, and the beam size ({FWHM of the corresponding convolved Gaussian beam (see Sect.~\ref{subsec:spectral_line_surveys})} for the ALMA lines{;} HPBW for the SD lines) of the observations are listed in the top right corner of each panel.}
    \label{fig:15194-5115_29SiO}
\end{figure*}
\plotfig[0.8\linewidth]{IRAS_15194-5115}{30SiO}{As in Fig.~\ref{fig:15194-5115_29SiO}, but for IRAS 15194-5115, $^{30}$SiO}\vspace{-15pt}

\plotfig[0.95\linewidth]{IRAS_15082-4808}{SiO}{As in Fig.~\ref{fig:15194-5115_29SiO}, but for IRAS 15082-4808, SiO}\vspace{-15pt}
\plotfig[0.95\linewidth]{IRAS_15082-4808}{29SiO}{As in Fig.~\ref{fig:15194-5115_29SiO}, but for IRAS 15082-4808, $^{29}$SiO}\vspace{-15pt}
\plotfig[0.325\linewidth]{IRAS_15082-4808}{30SiO}{As in Fig.~\ref{fig:15194-5115_29SiO}, but for IRAS 15082-4808, $^{30}$SiO}\vspace{-15pt}

\plotfig[0.95\linewidth]{IRAS_07454-7112}{SiO}{As in Fig.~\ref{fig:15194-5115_29SiO}, but for IRAS 07454-7112, SiO}\vspace{-15pt}
\plotfig[0.95\linewidth]{IRAS_07454-7112}{29SiO}{As in Fig.~\ref{fig:15194-5115_29SiO}, but for IRAS 07454-7112, $^{29}$SiO}\vspace{-15pt}
\plotfig[0.325\linewidth]{IRAS_07454-7112}{30SiO}{As in Fig.~\ref{fig:15194-5115_29SiO}, but for IRAS 07454-7112, $^{30}$SiO}\vspace{-15pt}

\plotfig[0.7\linewidth]{AFGL_3068}{SiO}{As in Fig.~\ref{fig:15194-5115_29SiO}, but for AFGL 3068, SiO}\vspace{-15pt}

\plotfig[0.95\linewidth]{IRC+10216}{SiO}{As in Fig.~\ref{fig:15194-5115_29SiO}, but for IRC+10216, SiO}\vspace{-15pt}
\plotfig[0.95\linewidth]{IRC+10216}{29SiO}{As in Fig.~\ref{fig:15194-5115_29SiO}, but for IRC+10216, $^{29}$SiO}\vspace{-15pt}
\plotfig[0.95\linewidth]{IRC+10216}{30SiO}{As in Fig.~\ref{fig:15194-5115_29SiO}, but for IRC+10216, $^{30}$SiO}\vspace{-15pt}


\plotfig[0.95\linewidth]{IRAS_15194-5115}{29SiS}{As in Fig.~\ref{fig:15194-5115_29SiO}, but for IRAS 15194-5115, $^{29}$SiS}\vspace{-15pt}
\plotfig[0.5\linewidth]{IRAS_15194-5115}{Si34S}{As in Fig.~\ref{fig:15194-5115_29SiO}, but for IRAS 15194-5115, Si$^{34}$S}\vspace{-15pt}

\plotfig[0.95\linewidth]{IRAS_15082-4808}{SiS}{As in Fig.~\ref{fig:15194-5115_29SiO}, but for IRAS 15082-4808, SiS}\vspace{-15pt}
\plotfig[0.95\linewidth]{IRAS_15082-4808}{29SiS}{As in Fig.~\ref{fig:15194-5115_29SiO}, but for IRAS 15082-4808, $^{29}$SiS}\vspace{-15pt}
\plotfig[0.325\linewidth]{IRAS_15082-4808}{30SiS}{As in Fig.~\ref{fig:15194-5115_29SiO}, but for IRAS 15082-4808, $^{30}$SiS}\vspace{-15pt}
\plotfig[0.875\linewidth]{IRAS_15082-4808}{Si34S}{As in Fig.~\ref{fig:15194-5115_29SiO}, but for IRAS 15082-4808, Si$^{34}$S}\vspace{-15pt}

\plotfig[0.95\linewidth]{IRAS_07454-7112}{SiS}{As in Fig.~\ref{fig:15194-5115_29SiO}, but for IRAS 07454-7112, SiS}\vspace{-15pt}
\plotfig[0.95\linewidth]{IRAS_07454-7112}{29SiS}{As in Fig.~\ref{fig:15194-5115_29SiO}, but for IRAS 07454-7112, $^{29}$SiS}\vspace{-15pt}
\plotfig[0.95\linewidth]{IRAS_07454-7112}{30SiS}{As in Fig.~\ref{fig:15194-5115_29SiO}, but for IRAS 07454-7112, $^{30}$SiS}\vspace{-15pt}
\plotfig[0.95\linewidth]{IRAS_07454-7112}{Si34S}{As in Fig.~\ref{fig:15194-5115_29SiO}, but for IRAS 07454-7112, Si$^{34}$S}\vspace{-15pt}

\plotfig[0.95\linewidth]{AFGL_3068}{SiS}{As in Fig.~\ref{fig:15194-5115_29SiO}, but for AFGL 3068, SiS}\vspace{-15pt}
\plotfig[0.875\linewidth]{AFGL_3068}{29SiS}{As in Fig.~\ref{fig:15194-5115_29SiO}, but for AFGL 3068, $^{29}$SiS}\vspace{-15pt}
\plotfig[0.8\linewidth]{AFGL_3068}{30SiS}{As in Fig.~\ref{fig:15194-5115_29SiO}, but for AFGL 3068, $^{30}$SiS}\vspace{-15pt}
\plotfig[0.8\linewidth]{AFGL_3068}{Si34S}{As in Fig.~\ref{fig:15194-5115_29SiO}, but for AFGL 3068, Si$^{34}$S}\vspace{-15pt}

\plotfig[0.95\linewidth]{IRC+10216}{SiS}{As in Fig.~\ref{fig:15194-5115_29SiO}, but for IRC+10216, SiS}\vspace{-15pt}
\plotfig[0.95\linewidth]{IRC+10216}{29SiS}{As in Fig.~\ref{fig:15194-5115_29SiO}, but for IRC+10216, $^{29}$SiS}\vspace{-15pt}
\plotfig[0.95\linewidth]{IRC+10216}{30SiS}{As in Fig.~\ref{fig:15194-5115_29SiO}, but for IRC+10216, $^{30}$SiS}\vspace{-15pt}
\begin{figure*}[h]
    \centering
    \includegraphics[width=0.95\linewidth]{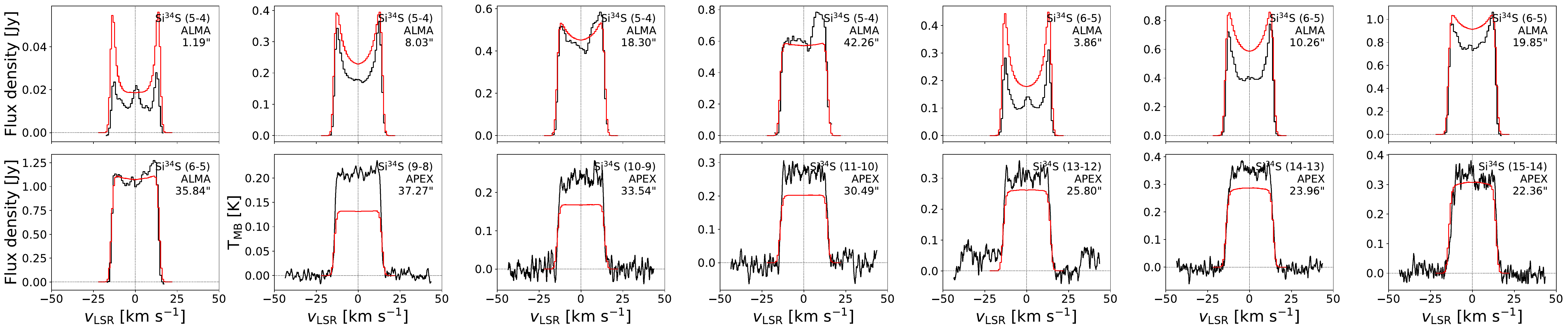}
    \caption{As in Fig.~\ref{fig:15194-5115_29SiO}, but for IRC+10216, Si$^{34}$S}
    \label{fig:10216_Si34S}
\end{figure*}


\end{appendix}

\end{document}